\documentclass{article}
\usepackage{times}
\usepackage{mathptmx}
\usepackage{epsfig}
\usepackage{algorithm}
\usepackage{algorithmic}

\usepackage{color}

\title{Noncooperatively Optimized Tolerance: Decentralized Strategic Optimization in Complex Systems}
\author{Yevgeniy Vorobeychik \quad Jackson R. Mayo\\ Robert C. Armstrong \quad Joseph R. Ruthruff\\
\normalsize{Sandia National Laboratories}\\
\normalsize{P.O. Box 969}\\
\normalsize{Livermore, CA 94551}
}

\date{}

\newcommand{\figw}{1.1in}
\newcommand{\figh}{\figw}
\newcommand{\figww}{2.25in}
\newcommand{\fighh}{1.7in}

\begin{document}


\maketitle

\begin{abstract}
We introduce noncooperatively optimized tolerance (NOT), a generalization of highly optimized tolerance (HOT) that involves strategic (game theoretic) interactions between parties in a complex system.
We illustrate our model in the forest fire (percolation) framework.
As the number of players increases, our model retains features of HOT, such as robustness, high yield combined with high density, and self-dissimilar landscapes, but also develops features of self-organized criticality (SOC) when the number of players is large enough.
For example, the forest landscape becomes increasingly homogeneous and protection from adverse events (lightning strikes) becomes less closely correlated with the spatial distribution of these events.
While HOT is a special case of our model, the resemblance to SOC is only partial; for example, the distribution of cascades, while becoming increasingly heavy-tailed as the number of players increases, also deviates more significantly from a power law in this regime.
Surprisingly, the system retains considerable robustness even as it becomes fractured, due in part to emergent cooperation between neighboring players.
At the same time, increasing homogeneity promotes resilience against changes in the lightning distribution, giving rise to intermediate regimes where the system is robust to a particular distribution of adverse events, yet not very fragile to changes.
\end{abstract}

\section{Introduction}

Highly optimized tolerance (HOT) and self-organized criticality (SOC) have received considerable attention as alternative explanations of emergent power-law cascade distributions~\cite{Bak87,Carlson99}.
The SOC model~\cite{Bak87,Clar96,Henley93} posits that systems can naturally arrive at criticality and power-law cascades, independently of initial conditions, by following simple rule-based processes.
Among the important features of SOC are (a) self-similarity and homogeneity of the landscape, (b) fractal structure of cascades, (c) a small power-law exponent (i.e., heavier tails), and (d) low density and low yield (e.g., in the context of the forest fire model, described below).
HOT~\cite{Carlson99,Carlson00a,Carlson00b,Newman02}, in contrast, models complex systems that emerge as a result of optimization in the face of persistent threats.
While SOC is motivated by largely mechanical processes, the motivation for HOT comes from evolutionary processes and deliberately engineered systems, such as the electric power grid.
The key features of HOT are (a) a highly structured, self-dissimilar landscape, (b) a high power-law exponent, and (c) high density and high yield~\cite{Carlson00b}.

HOT and SOC can be cleanly contrasted in the context of the forest fire (percolation) model~\cite{Carlson99,Henley93}.
The forest fire model features a grid, usually two-dimensional, with each cell being a potential site for a tree.
Intermittently, lightning strikes one of the cells according to some probability distribution.
If there is a tree in the cell, it is set to burn.
At that point, a cascade begins: fires spread recursively from cells that are burning to neighboring cells that contain trees, engulfing the entire connected component in which they begin.
The main distinction between HOT and SOC in the forest fire model is how they conceive the process of growing trees in the grid.
In the classical forest fire model (SOC) a tree sprouts in every empty cell with some fixed probability $p$.
The lightning strike distribution is usually conceived as being uniform.
As the process of tree growth interleaves with burnouts, the system reaches criticality, at which burnout cascades (equivalently, sizes of connected components) follow a power-law distribution.
At criticality, the tree landscape is homogeneous and self-similar, and burnouts follow a fractal pattern.
Additionally, both the fraction of cells with trees before lightning (the density) and after lightning (the yield) are relatively low at criticality.
In contrast, the HOT model conceives of a global optimizer charged with deciding the configuration of each cell (i.e., whether a tree will grow or not).
What emerges globally as a consequence is a collection of large connected components of trees separated by ``barriers'' of no trees that prevent fires from spreading outside components.
This pattern, which is self-dissimilar, adapts to the specific spatial distribution of lightning strikes.
In certain cases, for example, when the lightning distribution is exponential or Gaussian, out of this adaptation emerges a precise balance of connected component sizes and fire probabilities so as to yield a power-law distribution of cascades (burnouts), with a higher exponent than the distribution of cascades in the SOC model.
The HOT model features both high yield and high density, as it is deliberately robust to lightning strikes with the specified distribution; however, it is also extremely fragile to changes in the lightning distribution, whereas SOC does not exhibit such fragility.
The HOT landscape tends to have a highly non-uniform distribution of ``fire breaks'', or areas where no trees are planted, whereas the SOC landscape is homogeneous.

A natural criticism of the HOT paradigm is that, in complex systems, it is difficult to conceive of a single designer that manages to optimally design such a system.
As a partial response, much work demonstrates that HOT yields qualitatively similar results when heuristic optimization or an evolutionary process is used~\cite{Carlson00a,Zhou02,Zhou05}.
Still, most complex systems, particularly those that are engineered, are not merely difficult to design globally, but are actually \emph{decentralized}, with many entities responsible for parts (often small) of the whole system.
Each entity is generally not motivated by global concerns, but is instead responding to local incentives, which may or may not align with global goals.
For example, the Internet is fundamentally a combination of autonomous entities, each making its own decisions about network topology, protocols, and composition, with different decisions made at different levels of granularity (some by large Internet service providers, some by large organizations connected to the Internet, some by small organizations and individual users).
Likewise, the electric grid emerges as a by-product of complex interactions among many self-interested parties, including electric utilities, electricity users (which are themselves businesses or individuals), and various government and regulatory entities that have their own interests in mind.
Most moderately complex engineered products are manufactured from components produced by different firms, each with its own goals driven primarily by the market within which it competes, and many of these components are further broken down and produced by their own set of suppliers, and so on.

Our central contribution is to model complex systems as complex patterns of strategic interactions among self-interested players making independent decisions.
We conceive that out of \emph{strategic interactions} of such self-interested players emerges a system that is optimized \emph{jointly} by all players, rather than \emph{globally} by a single ``engineer''.
Thus, we call our model \emph{noncooperatively optimized tolerance} (NOT).
Formally, our model is game theoretic, and we seek to characterize emergent properties of the system in a Nash equilibrium~\cite{Osborne94}.
Our model strictly generalizes the HOT framework, with HOT being the special case of a game with a single player.

\section{A Game Theoretic Forest Fire Model}

We begin by introducing some general game theoretic notions, and then instantiate them in the context of a forest fire model.
A \emph{game} is described by a set of players $I$, numbering $m = |I|$ in all, where each player $i \in I$ chooses actions from a strategy set $S_i$ so as to maximize his \emph{utility} $u_i(\cdot)$.
Notably, each player's utility function depends on the actions of other players as well as his own, and so we denote by $u_i(s) = u_i(s_i,s_{-i})$ the utility to player $i$ when he plays a strategy $s_i$ and others jointly play $s_{-i} \equiv (s_1,\ldots,s_{i-1},s_{i+1},\ldots,s_m)$, where these combine to form a joint strategy profile $s = (s_1,\ldots,s_i,\ldots,s_m)$.
In our context, each player controls a portion of a complex system and is responsible for engineering his ``domain of influence'' against perceived threats, just as in the HOT model.
The distinction with the HOT model is that the interests of different players may be opposed if, say, an action that is desirable for one has a negative impact on another (for example, one player may dump his trash on another's territory).
Such interdependencies are commonly referred to as \emph{externalities}~\cite{Mascolell95}, and form a central aspect of our model.
However, HOT arises as a special case of our construction, when the game has a single player.

We implement the game theoretic conception of complex system engineering in the familiar two-dimensional forest fire model, thereby allowing direct contrast with the now mature literature on HOT and SOC\@.
In the NOT forest fire model, each player is allotted a portion of the square grid over which he optimizes his yield less cost of planting trees.\footnote{We note the resemblance of our grid division into subplots to the framework studied by Kauffman et al.~\cite{Kauffman94}, which divides a lattice in a similar manner, but with the goal of studying joint optimization of a global objective, rather than strategic interactions among players controlling different plots and having different goals.}
Let $G_i$ be the set of grid cells under player $i$'s direct control, let $s_i$ be player $i$'s strategy expressed as a vector in which $s_{i,g} = 1$ if $i$ plants a tree in grid cell $g$ and $s_{i,g} = 0$ otherwise, and let $\Pr\{\,g = 1 \mid s, s_{i,g} = 1\,\}$ be the probability (with respect to the lightning distribution) that a tree planted in cell $g$ survives a fire given the joint strategy (planting) choices of all players.
Since exactly one player controls each grid cell, we simplify notation and use $s_g = s_{i,g}$ where $i$ is the player controlling grid cell $g$.
Let $N_i = |G_i|$ be the number of grid cells under $i$'s control and $\rho_i$ be the density of trees planted by $i$,
\[
\rho_i = \frac{1}{N_i} \sum_{g \in G_i} s_g.
\]
Define the yield for player $i$ to be
\[
Y_i(s) = \sum_{g \in G_i} \Pr\{\,g = 1 \mid s\,\}s_{g}
\]
(it is convenient to define the yield as an absolute number of trees).
We assume further that each tree planted by a player incurs a fixed cost $c$.
The utility of player $i$ is then
\[
u_i(s) = \sum_{g \in G_i} (\Pr\{\,g = 1 \mid s\,\} - c)s_{i,g}  = Y_i(s) - cN_i\rho_i.
\]

The result of joint decisions by all players is a grid that is partially filled by trees, with overall density $\rho(s)$ and overall yield $Y(s)$ given by a sum ranging over the entire grid $G$, i.e., $Y(s) = \sum_{g \in G} \Pr\{\,g = 1 \mid s\,\}s_{g}$.
Let $N$ be the number of cells in the entire grid.
We then define \emph{global utility (welfare)} as
\[
W(s) = \sum_{i \in I} u_i(s) = Y(s) - cN\rho(s).
\]
Note that when $m=1$, $W(s)$ coincides with the lone player's utility.
A part of our endeavor below is to characterize $W(s^*)$ and $\rho(s^*)$ when $s^*$ is a Nash equilibrium, defined as a configuration of joint decisions by all players such that no individual player can gain by choosing an alternative strategy $s_i'$ (alternative configuration of trees planted) \emph{keeping the decisions of other players $s_{-i}^*$ fixed}.

We systematically vary several parameters of the model.
The first, which is the main subject of this work, is the number of players $m$.
Fixing the size of the grid at $N = 128 \times 128$,\footnote{This was the largest grid size on which we could approximate equilibria in reasonable time.} we vary the number of players between the two extremes, from $m = 1$ to $m = N$.
The former extreme corresponds precisely to the HOT setting, while in the latter the players are entirely myopic in their decision problems, each concerned with only a single cell of the grid.
The negative externalities of player decisions are clearly strongest in the latter case.
The entire range of player variation is $m \in \{1,2^2,4^2,8^2,16^2,32^2,64^2,128^2\}$.
The second parameter that we vary is the cost of planting trees: $c \in \{0, 0.25, 0.5, 0.75, 0.9\}$.
Finally, we vary the scale of the lightning distribution, which is always a truncated Gaussian centered at the top left corner of the grid.
We let the variance (of the Gaussian before truncation) be $N / v$, and vary $v \in \{0.1,1,10,100\}$.
For example, at $v = 1$ the standard deviation of the Gaussian covers, roughly, the entire size of the grid, and at $v = 0.1$ the distribution of lightning strikes is approximately uniform over the grid.
In contrast, $v = 100$ gives a distribution with lightning strikes highly concentrated in the top left corner of the grid.

The question yet to be addressed is how to partition the grid into regions of influence for a given number of players $m$.
We do this in the most natural way by partitioning the grid into $m$ identical square subgrids, ensuring throughout that $m$ is a power of 4.

\section{Analysis of the NOT Forest Fire Model}

To build some intuition about our model, consider first a one-dimensional forest fire setting.
Since in one dimension sequences of planted cells (1's) are interleaved with unplanted sequences (0's), we define $k$ to be the length of a planted sequence and $l$ be the length of an unplated sequence, and assume that $1 \ll k \ll N$.
First, consider the case with $m=1$ and assume that $c < 1 - 1/N$.
We further assume that $k$ is identical for all sequences of 1's (when $k \ll N$, this is almost with no loss of generality, since 1's can be swapped, keeping the density constant, without changing the utility) and note that in an optimal solution $l=1$.
The utility of the player (and global utility) is then
\[
u_i(k) = W(k) = N\rho(k)\left(1 - \frac{k}{N} - c\right),0
\]
where $\rho(k) = k/(k+1)$.
If we view $k$ as a continuous variable, we can obtain a maximizer, $k^* = O(\sqrt{N(1-c)})$.
In contrast, if we consider the case with each player occupying a single grid cell (i.e., $m=N$), $k^E = O(N(1-c))$.
While the density of planting approaches 1 in both the optimal and equilibrium configurations as $N$ increases (as long as $N(1-c) \gg 1$), it turns out that the equilibrium density is generally higher than optimal (all the results discussed here are derived in supporting online material).
This agrees with our intuition on the consequence of negative externalities of decentralized planting decisions: when a player decides whether to plant a tree, he takes into account only the concomitant chance of his own tree burning down, and not the global impact the decision has on the sizes of cascades.

Armed with some intuition based on the one-dimensional model, we now turn to our main subject, the two-dimensional forest fire model---varying systematically the number of players, planting cost, and lightning distribution as described above.
A full analysis of the two-dimensional model in all the relevant parameters is beyond mathematical tractability.
Furthermore, the problem of computing exact equilibria, or even exact \emph{optima} for any player, is intractable, as the size of the space of joint player strategies in our setting is $2^{16384}$ (for example, at one extreme, we need to compute or approximate an equilibrium in a game with 16384 players, each having binary strategies).

Despite the daunting size of the problem, it turns out that simple iterative algorithms for approximating equilibria as well as optimal decisions by individual players are extremely effective.
Specifically, we use the following procedure for approximating Nash equilibria, building on previous methodological work in simulation-based game theoretic analysis~\cite{Sureka05,Wellman05,Seale06,Vorobeychik08a,Vorobeychik08b,Vorobeychik09}:
\begin{enumerate}
\item Start with no trees planted as the initial joint strategy profile $s$
\item For a fixed number of iterations:
  \begin{enumerate}
    \item For each player $i$:
      \begin{enumerate}
        \item Fix the decisions of other players $s_{-i}$
        \item With probability $p$, compute $\hat{s}_i \leftarrow \mathrm{OPT}(s_{-i})$, the optimal decision of player $i$ given $s_{-i}$; with probability $1-p$, let $\hat{s}_i \leftarrow s_i$
        \item Set $s_i \leftarrow \hat{s}_i$
      \end{enumerate}
  \end{enumerate}
\end{enumerate}
This procedure is a variant of \emph{best response dynamics}, which is a well-known simple heuristic for learning in games~\cite{Fudenberg98}.
While convergence properties of this heuristic are somewhat weak~\cite{Fudenberg98,Vorobeychik08a}, it has proved to be quite effective at approximating equilibria computationally~\cite{Sureka05,Vorobeychik08a}, and has the additional feature of being a principled model for adaptive behavior of goal-driven agents in a strategic setting~\cite{Fudenberg98}.

The inner loop of the algorithm involves computing an optimal (best) response of a player $i$, which we already noted is in general intractable.
Carlson and Doyle~\cite{Carlson00a} used a deterministic greedy heuristic to compute an optimum over an entire grid of size $64 \times 64$, starting with a grid devoid of trees and iteratively choosing a grid cell that maximizes global utility given the planting choices from the previous iterations.
However, we use a larger ($128 \times 128$) grid, and additionally must run the optimization heuristic multiple times as an inner loop of equilibrium approximation, so their approach is too computationally intensive to be practical in our setting.
Instead, we utilize the somewhat lighter-weight method of \emph{sampled fictitious play}~\cite{Lambert05,Epelman11}, which allows us to more finely control the tradeoff between the amount of optimization search and the incremental impact of additional search on solution quality.
In sampled fictitious play, each grid cell controlled by player $i$ becomes a ``player'' in a cooperative subgame (where each cell has $i$'s utility as its goal), and random subsets of cells are iteratively chosen to make simultaneous optimizing decisions given a uniform random sample of choices by the rest of the grid from a fixed window of previous iterations.
Random \emph{exploration} is introduced by occasionally replacing historical actions of ``players'' (cells controlled by $i$) with randomly chosen actions.
In our implementation, it turned out to be most effective to let the history window size be 1, which makes sampled fictitious play resemble myopic best response dynamics.
Since each grid cell has only two actions, we choose the myopically best action, determined by the size of the connected component of trees to which the cell belongs.

Insofar as the results below are the outcomes of the above algorithm, they represent, approximately, principled predictions of coevolution of goal-directed players whose incentives may not align with the global objective.
As such, our simulation results have an additional advantage over the 1-D mathematical characterization, which does not allow direct insight into \emph{which} of the many possible equilibrium configurations is likely to be reached by adaptive players.

\subsection{Global Utility}

\begin{figure}[ht]
\centering
\begin{tabular}{cc}
\includegraphics[width=\figww]{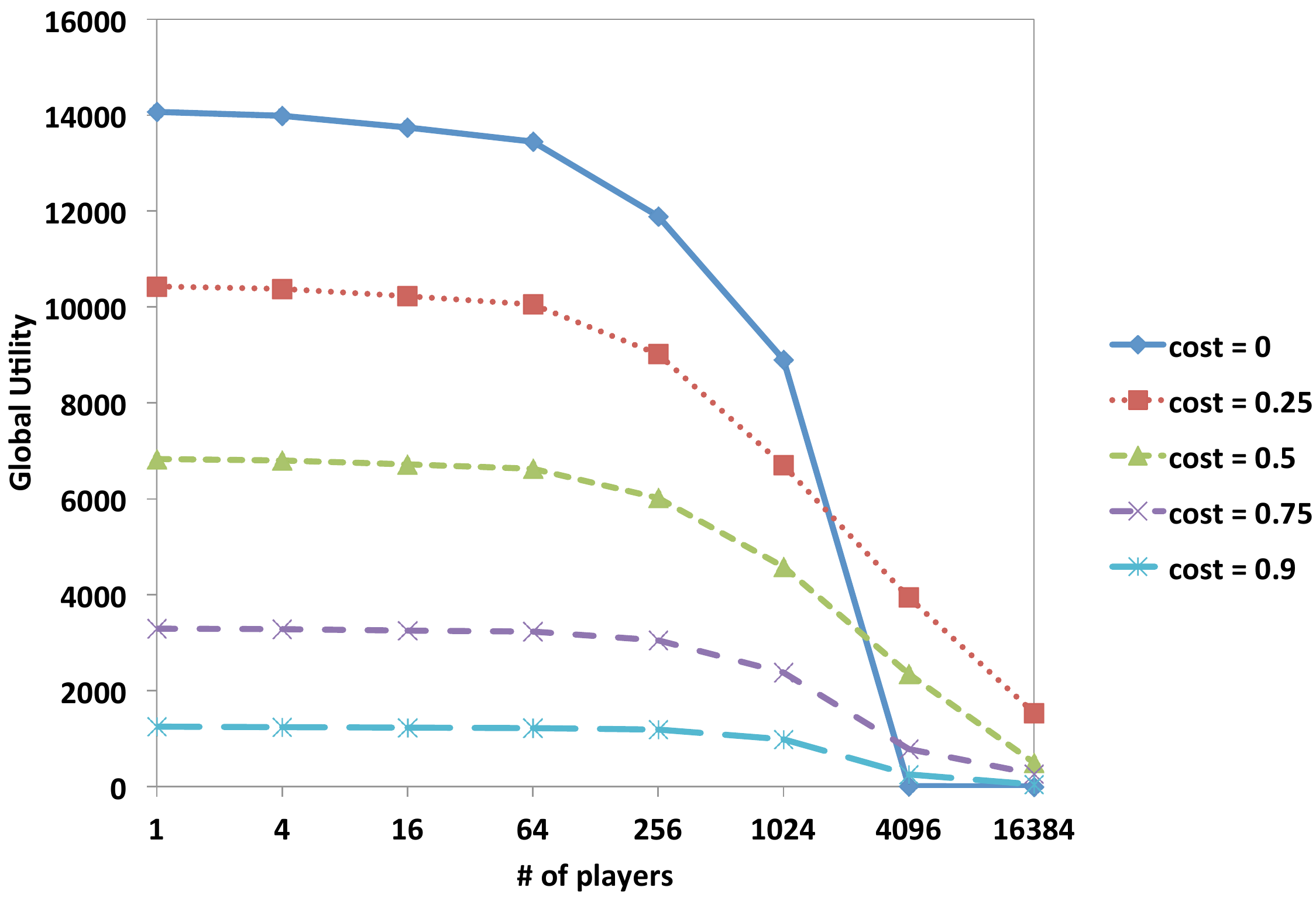}
&
\includegraphics[width=\figww]{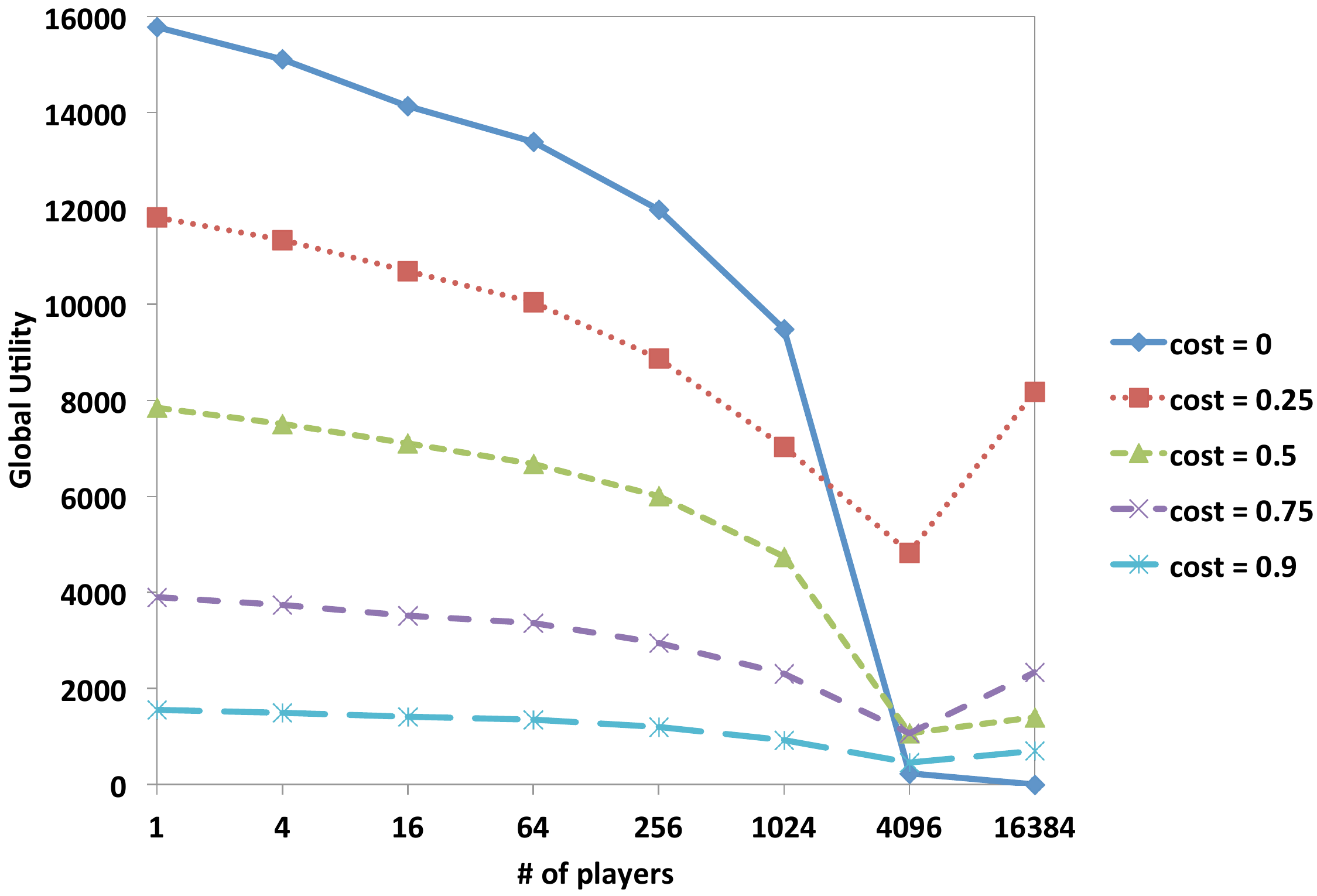} 
\end{tabular}
\caption{Global utility $W(s^*)$ as a function of $m$ for $c \in \{0,0.25,0.5,0.75,0.9\}$.  Left: $v=0.1$ (nearly uniform distribution).  Right: $v=100$ (highly concentrated distribution).}
\label{F:utility}
\end{figure}

Our first question concerns the variation of global utility $W(s^*)$ with the number of players $m$, the cost $c$, and the parameter $v$ governing variance of the Gaussian lightning distribution.
First, note that $W(s^*)$ will be no better than optimal for $m>1$, and it seems intuitive that it is a non-increasing function of $m$.
Additionally, when $c=0$ and $m=N$, we anticipate a global utility of 0, since the only equilibria involve either all players or all but one planting trees (we argue this formally in the online supplement).
The questions we address next are: what happens when $1 < m < N$ and when $c > 0$?
Figure~\ref{F:utility} provides some answers.
First, when $c=0$, we notice that the initial drop in global utility is quite shallow for $m < 256$, particularly when the lightning distribution is relatively diffuse ($v < 100$).
However, once the number of players is relatively large, global utility drops dramatically, and nearly reaches 0 already when $m=4096$, even though this does not directly follow from an argument above.
For $c > 0$, the dropoff in global utility with the number of players becomes less dramatic.

\subsection{Density and Fire Break Distribution}

Our next task is to consider how the density changes with our parameters of interest.
Based on the observation above, we expect the density to be $1$, or nearly so, when $c=0$ and $m=N$.
The density should be appreciably below $1$ when $m=1$.
Furthermore, the density should decrease with increasing cost $c$.
In general, our intuition, based on all previous analysis, would suggest that density should increase with the number of players: after all, each player's decision to plant a tree does not account for the negative impact it has on other players.

\begin{figure}[ht]
\centering
\begin{tabular}{cc}
\includegraphics[width=\figww]{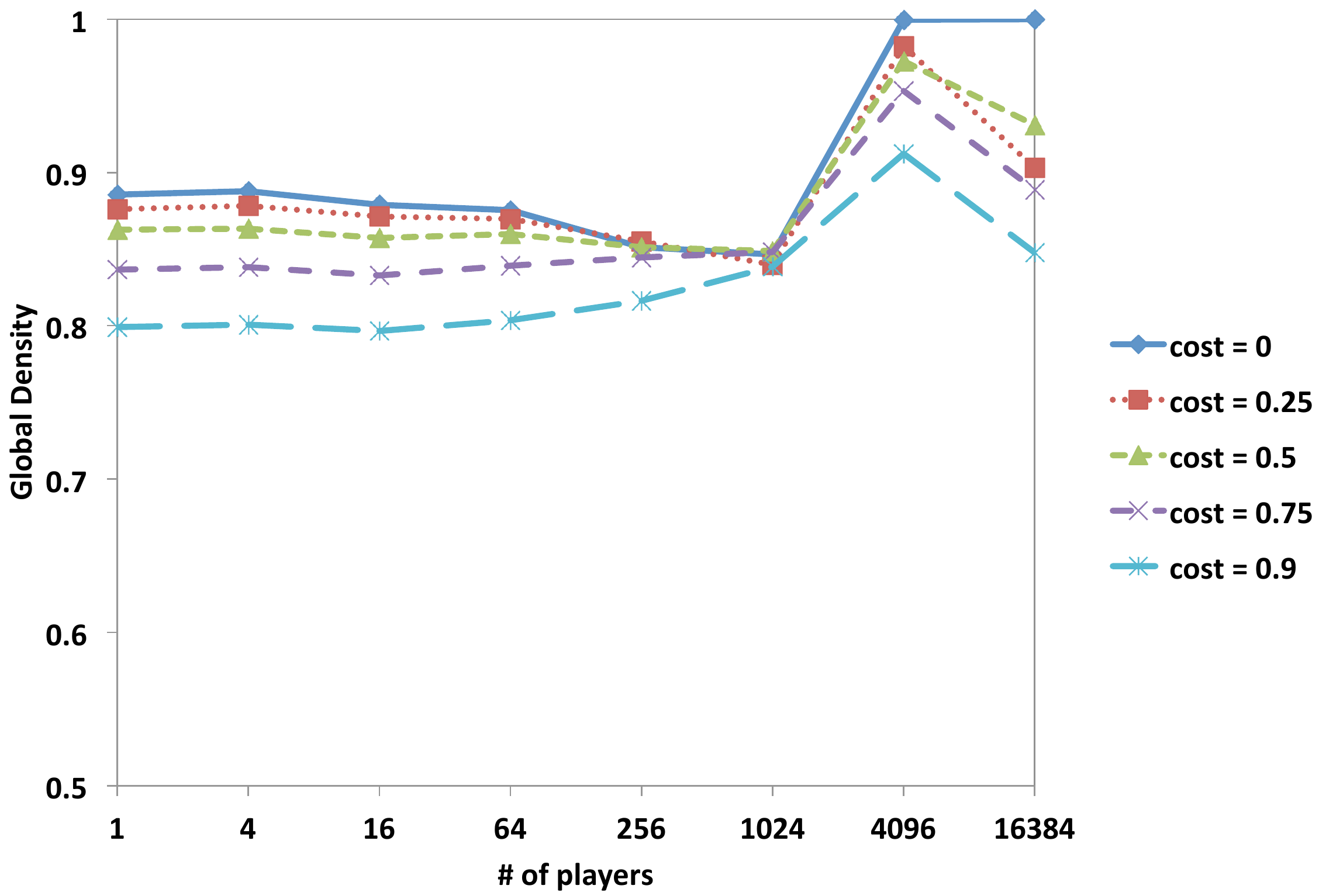}
&
\includegraphics[width=\figww]{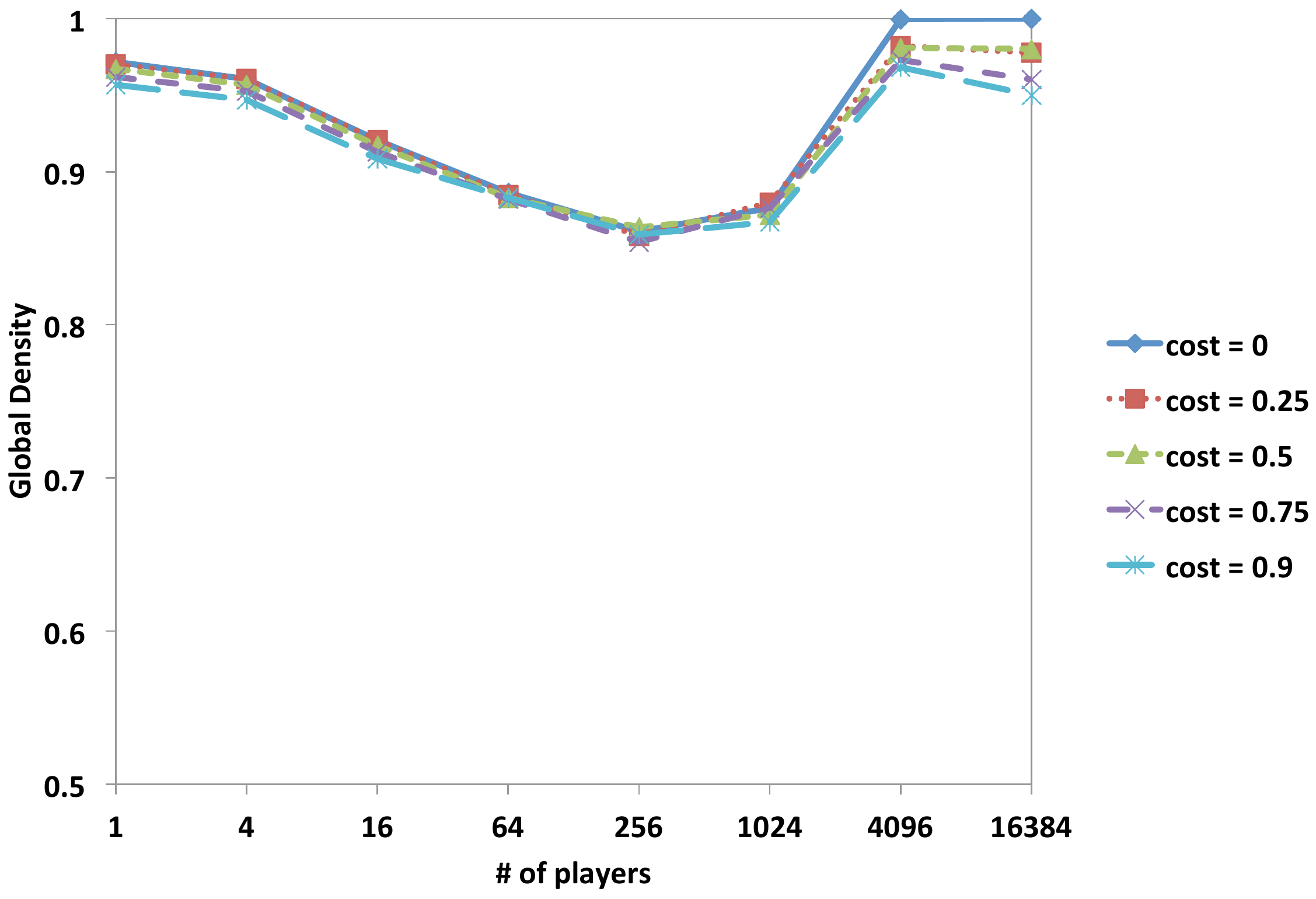} 
\end{tabular}
\caption{Density $\rho$ as a function of $m$ for $c \in \{0,0.25,0.5,0.75,0.9\}$.  Left: $v=0.1$ (nearly uniform distribution).  Right: $v=100$ (highly concentrated distribution).}
\label{F:density}
\end{figure}

\begin{figure}[ht]
\centering
\begin{tabular}{cccc}
\includegraphics[width=\figw,height=\figh]{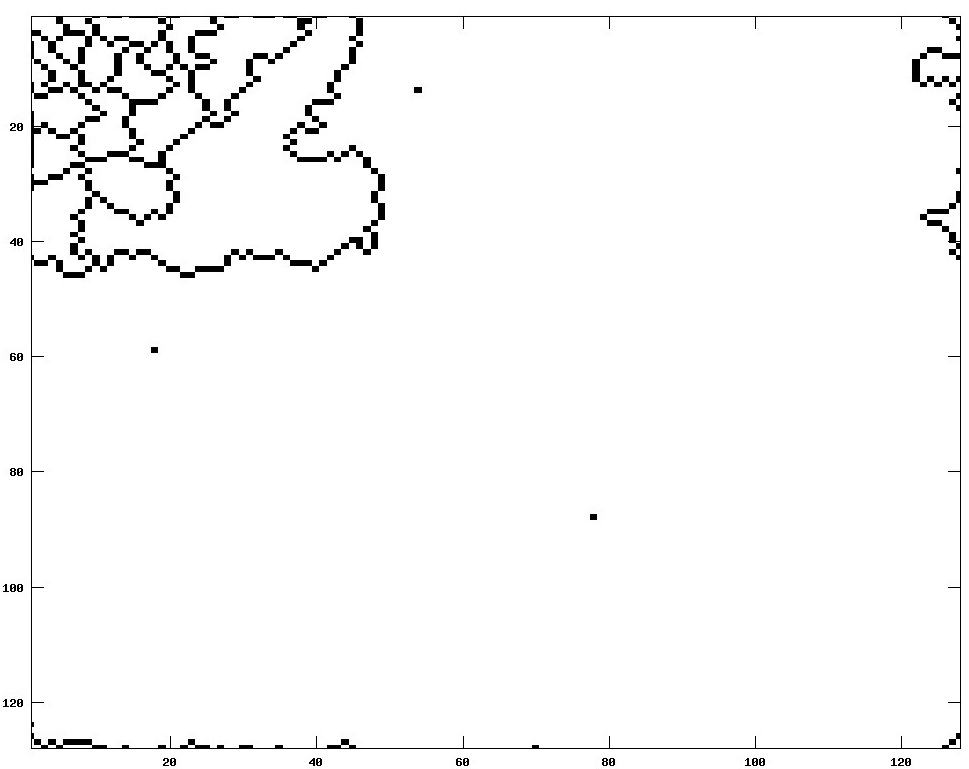}
& 
\includegraphics[width=\figw,height=\figh]{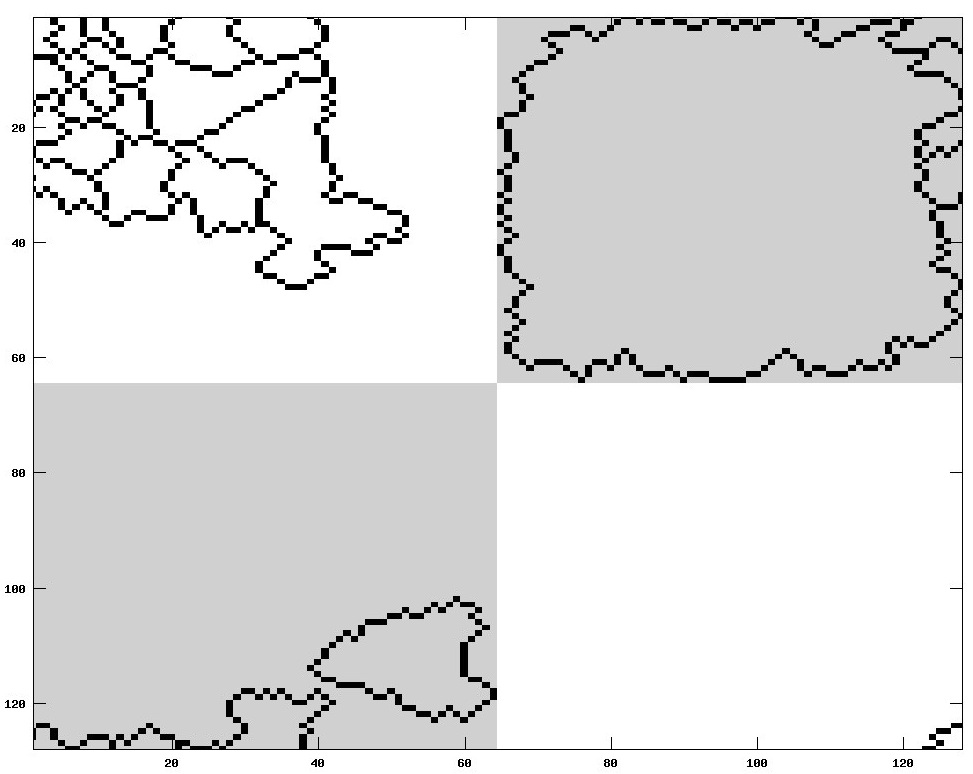}
&
\includegraphics[width=\figw,height=\figh]{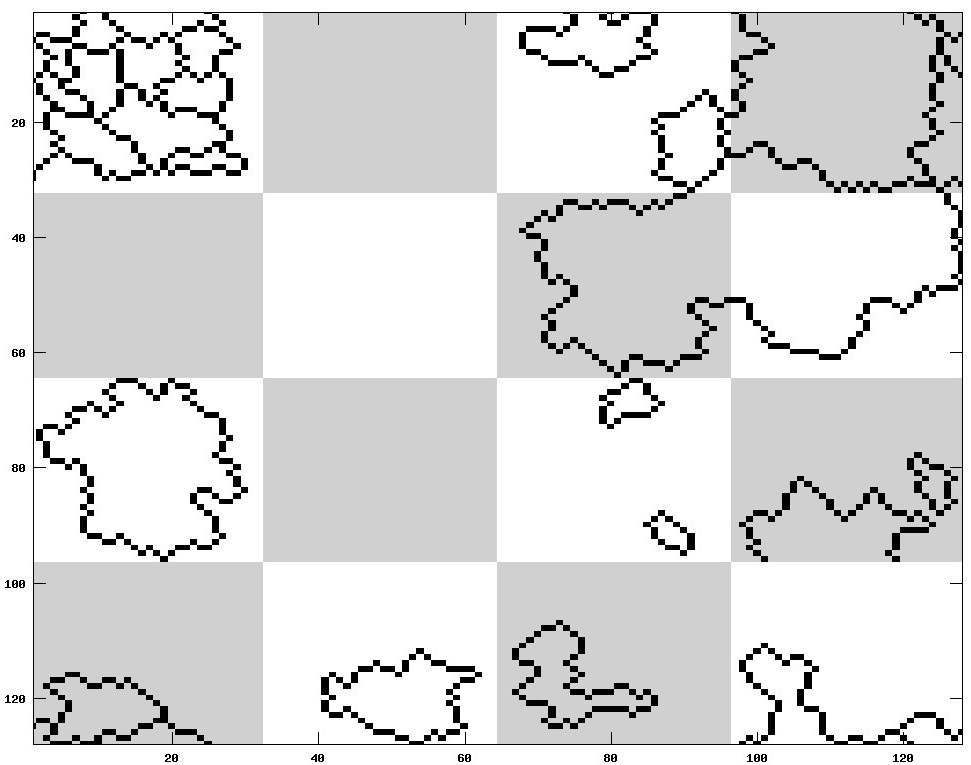}
&
\includegraphics[width=\figw,height=\figh]{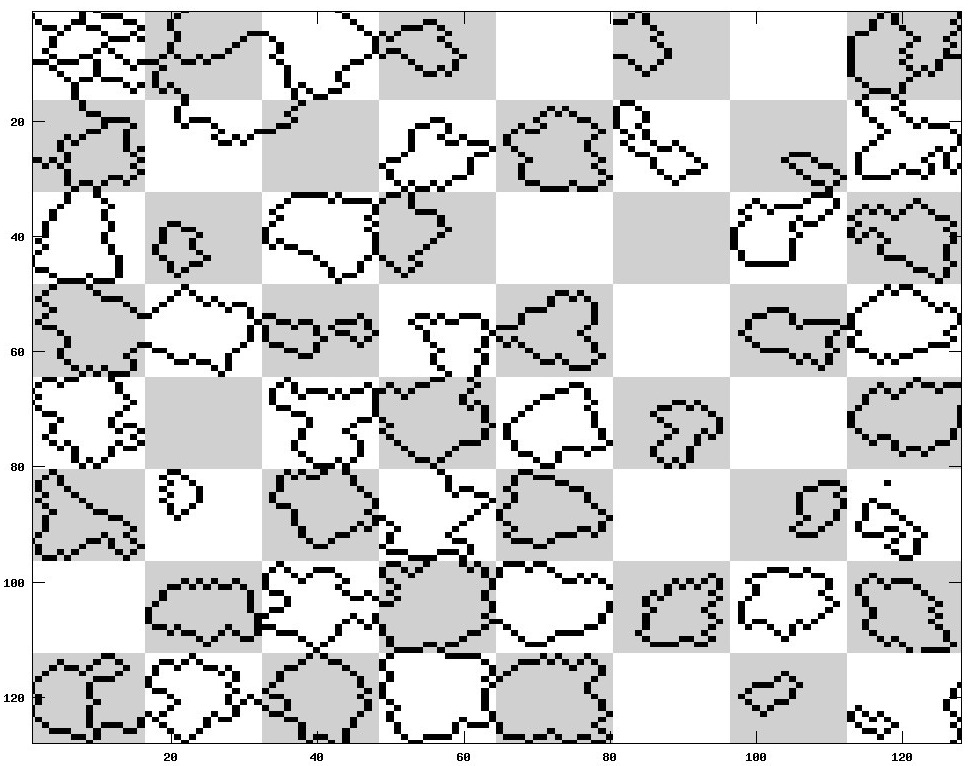}\\
(a) & (b) & (c) & (d) \\
\\
\includegraphics[width=\figw,height=\figh]{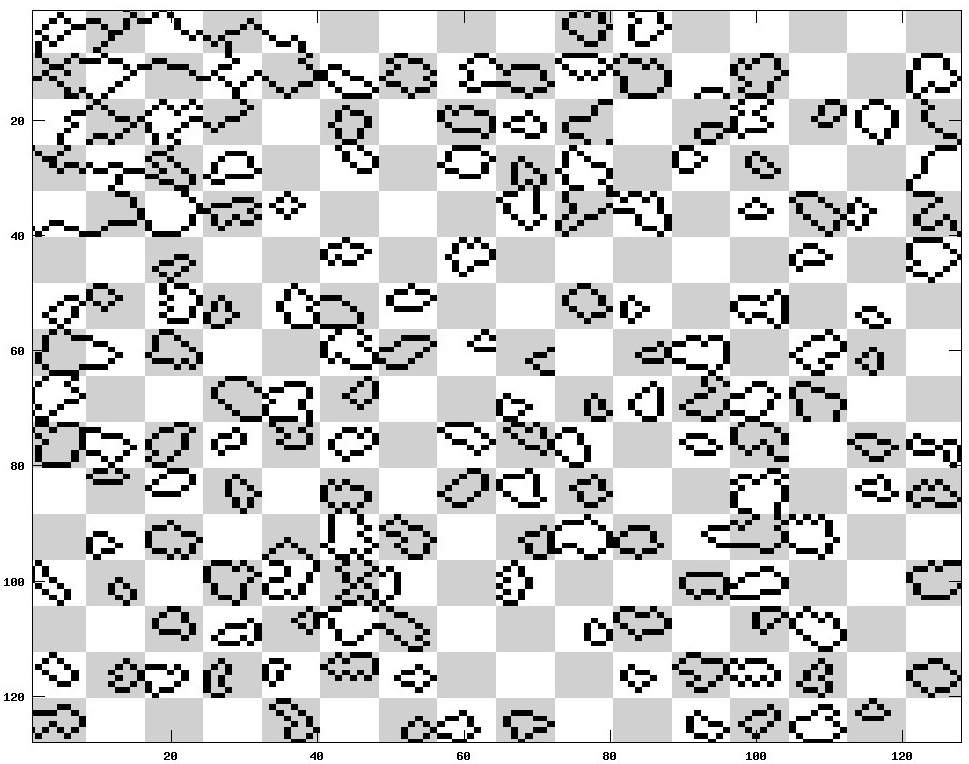}
&
\includegraphics[width=\figw,height=\figh]{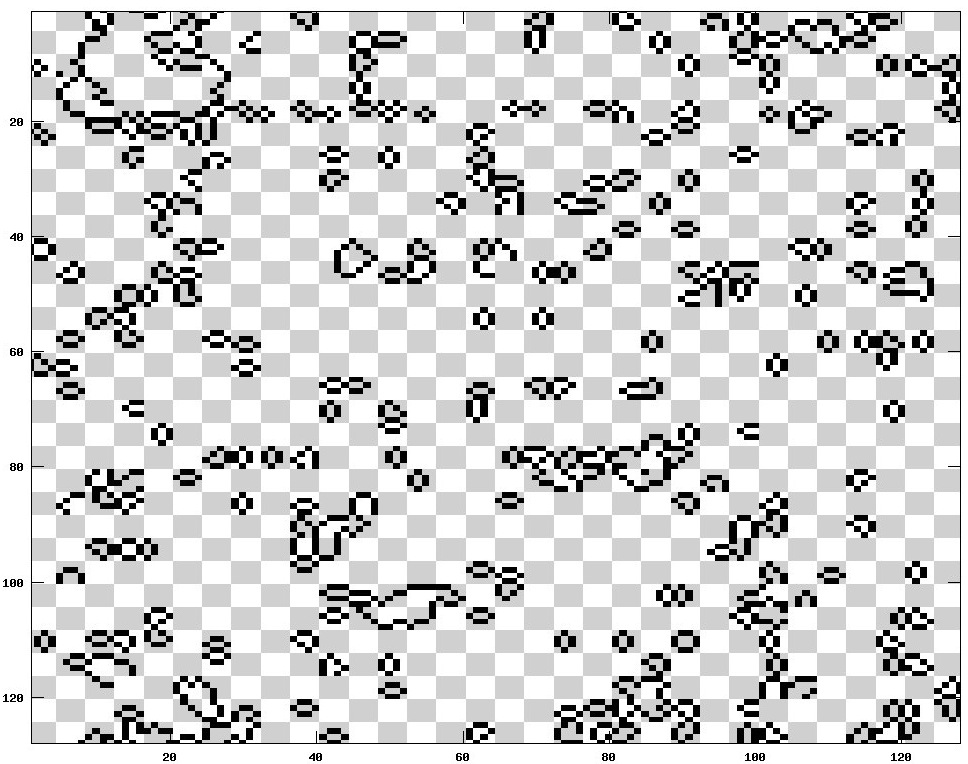}
&
\includegraphics[width=\figw,height=\figh]{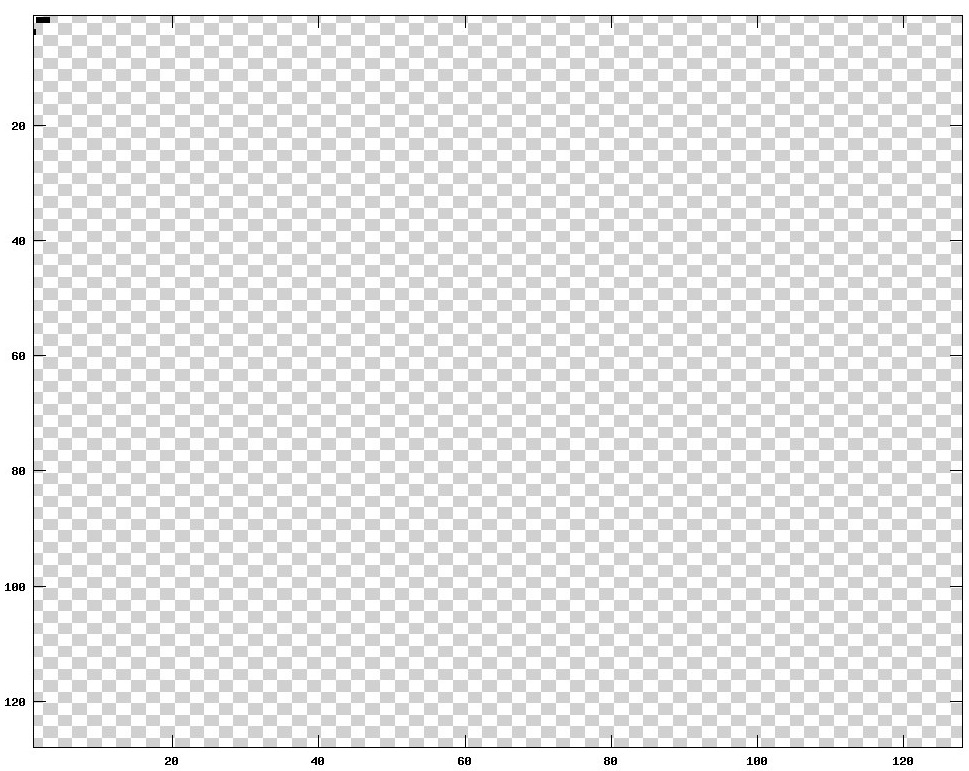} 
&
\includegraphics[width=\figw,height=\figh]{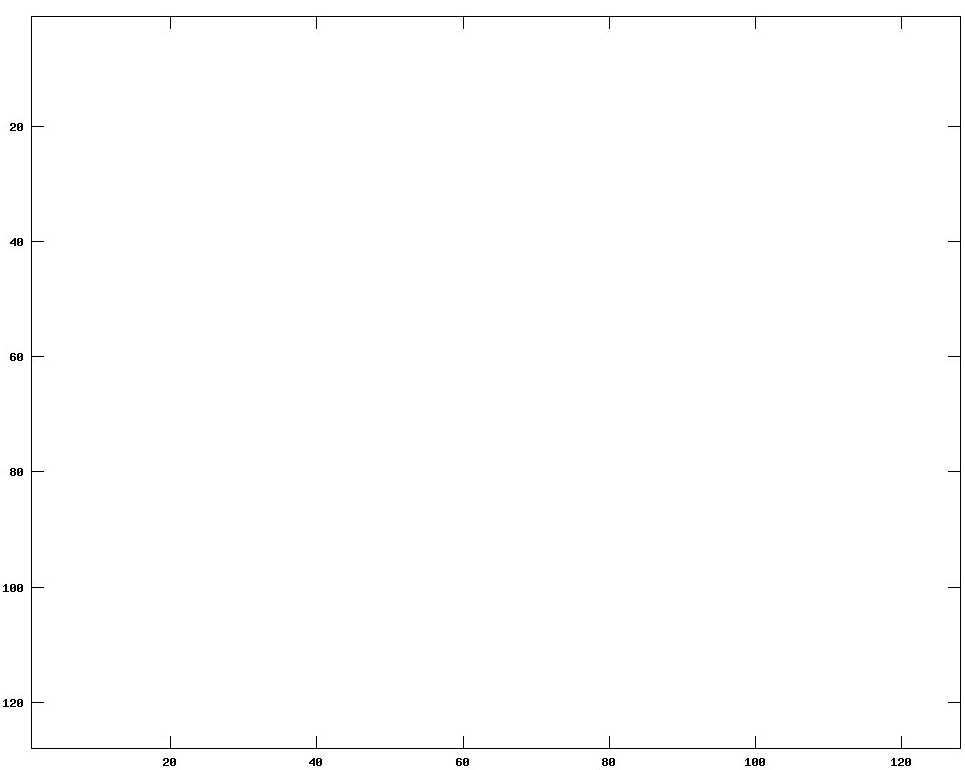}\\ 
(e) & (f) & (g) & (h)
\end{tabular}
\caption{Sample equilibrium grid configurations with $c=0$, $v=100$, and the number of players varied between $1$ and $N = 16384$. Blank cells are planted and marked cells are unplanted. Player domains of influence are shaded in a checkerboard pattern.  (a) $1$ player, equivalent to HOT; (b) $4$ players; (c) $16$ players; (d) $64$ players; (e) $256$ players; (f) $1024$ players; (g) $4096$ players; (h) $16384$ players.  To avoid clutter, we omit the checkerboard pattern with $N$ players.  Here, the grid is blank, which indicates that every grid cell contains a tree.}
\label{F:grid_c0_v100}
\end{figure}

Working from this intuition, the results in Figure~\ref{F:density} are highly counterintuitive: the overall density \emph{falls} with increasing number of players until $m$ reaches $1024$, and only when the number of players is very high ($4096$ and $N$) is it generally higher than the optimal density.
This dip is especially apparent for a highly concentrated lightning distribution ($v=100$).
To understand this phenomenon we must refer to Figure~\ref{F:grid_c0_v100}, showing actual (approximate) equilibrium grid configurations for varying numbers of players when $c=0$ and $v=100$.
We can observe that each player's myopic self-interest induces him to construct \emph{fire breaks} in his territory where none exist in a globally superior single-player configuration.
Thus, for example, contrast Figure~\ref{F:grid_c0_v100} (a) and (b).
In the former, most of the grid is filled with trees, and much of the action happens in the upper left corner (the epicenter of the lightning distribution), which is filled with fire breaks that confine fires to relatively small fractions of the grid.
In the latter, the upper left corner is now under the control of a single player, and other players find it beneficial to plant fire breaks of their own, since the ``wasted'' land amounts to only a small fraction of their landmass, and offers some protection against fire spread to the protected areas from ``poorly'' protected neighboring territories.
With more players, we see coordination between neighbors emerge, as they jointly build mutually beneficial fire breaks, but such cooperation is not global, and becomes increasingly diffuse with greater number of players.
Nevertheless, increasing the number of players results in a greater amount of total territory devoted to fire breaks by individual players or small local neighborhoods, and, as a result, an overall loss in planting density, which have observed in Figure~\ref{F:density}.

\begin{figure}[ht]
\centering
\begin{tabular}{cc}
\includegraphics[width=\figww]{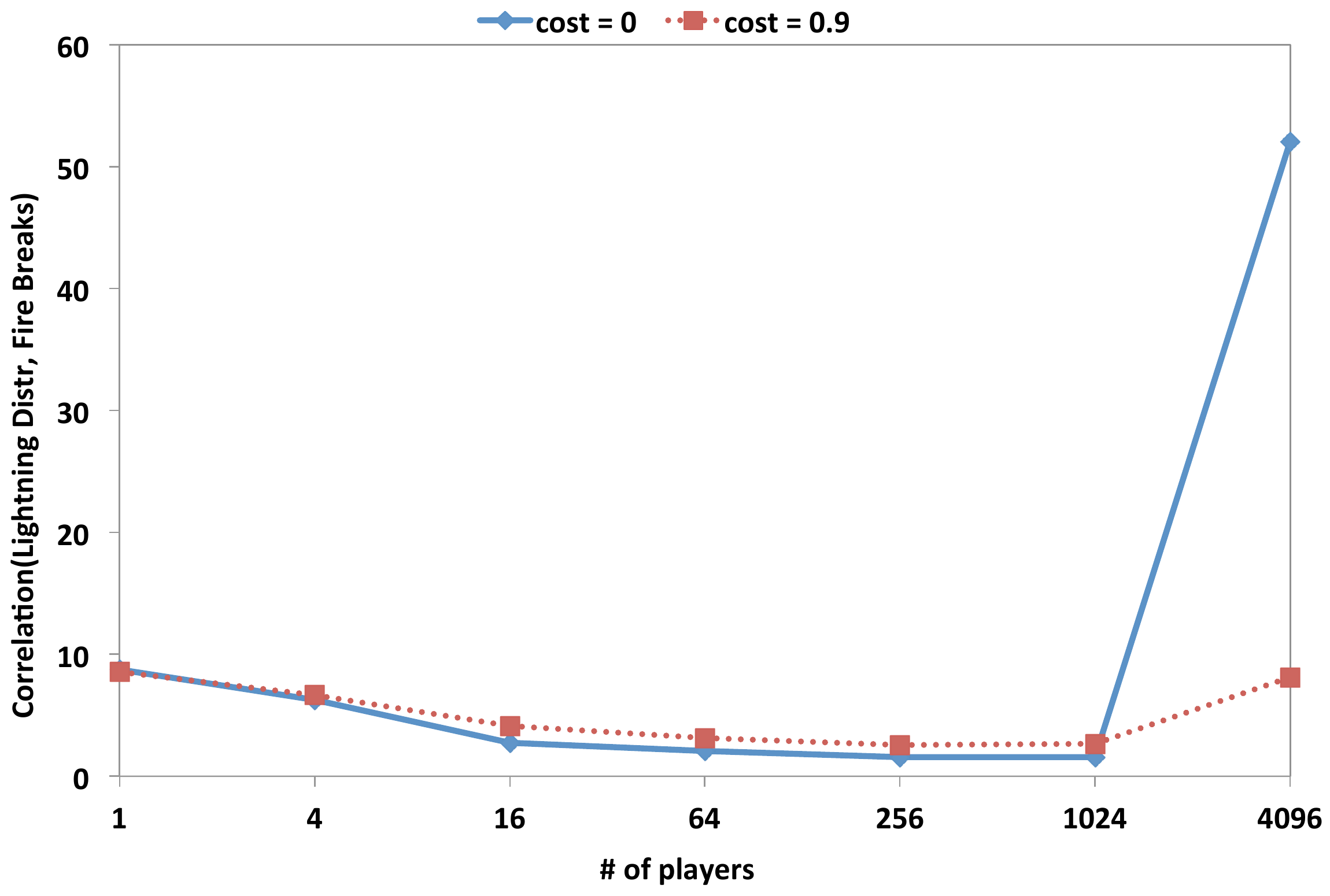} &
\includegraphics[width=\figww,height=\fighh]{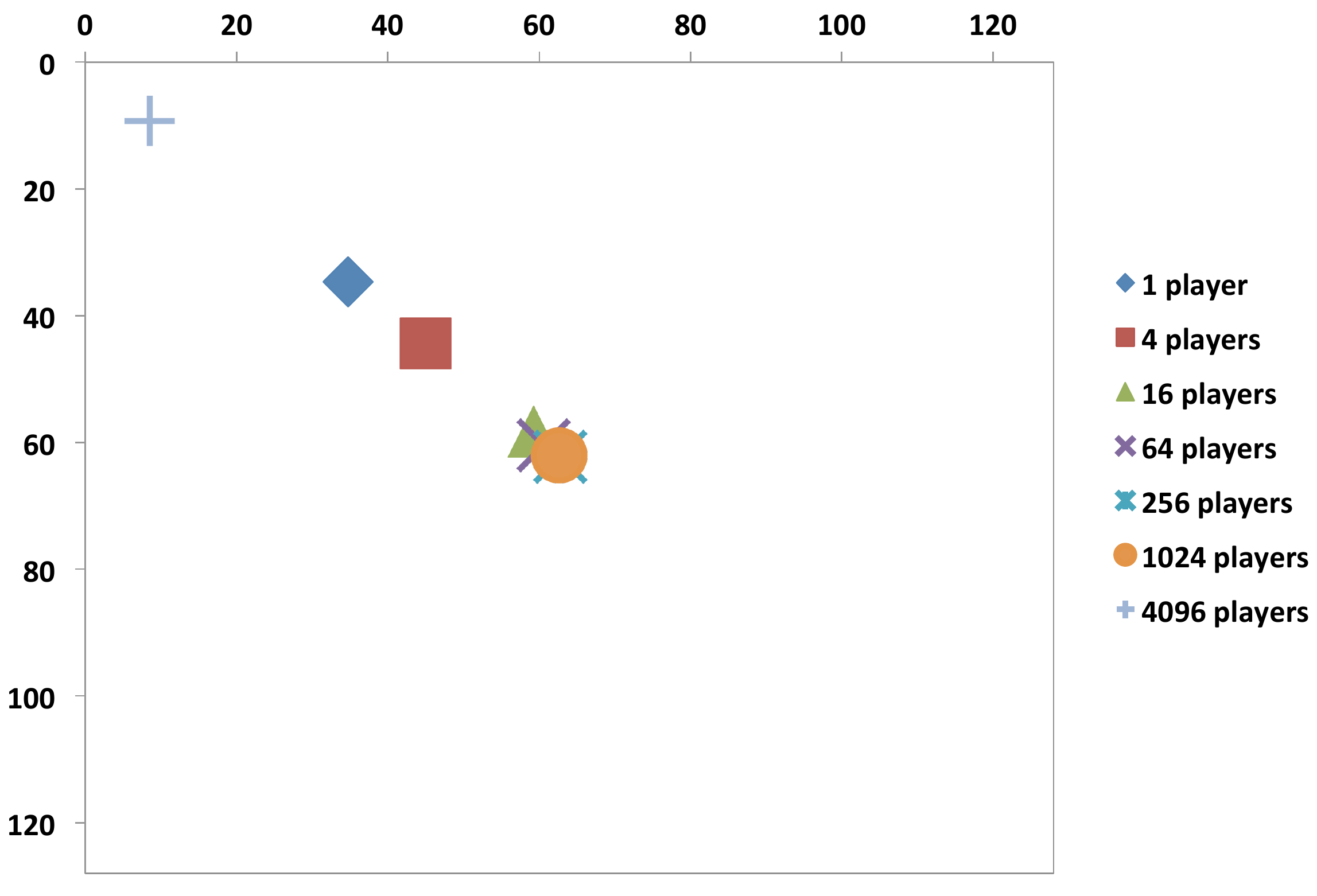}\\
\end{tabular}
\caption{Left: a measure of correlation ($C$, defined in the text) between the lightning distribution and the fire breaks (empty cells) across subgrids for $c=0$ and $c=0.9$.  As $C$ approaches 1, the locations of empty cells become essentially unrelated to the distribution of lightning strikes.  Right: centroid coordinates of the empty grid cells when $c=0$ (the results are similar when $c=0.9$).}
\label{F:fire_breaks}
\end{figure}

Since the density is decreasing for intermediate numbers of players, a natural hypothesis is that the fire breaks are distributed suboptimally.
We can observe this visually in Figure~\ref{F:grid_c0_v100}.
Specifically, the equilibrium grid configurations suggest that the location of fire breaks becomes less related to the lightning distribution as the number of players grows.
To measure this formally, we compute 
\[
C = \frac{\sum_{g \in G} p_g (1-s_g)}{1-\rho}.
\]
The numerator is the probability that lightning strikes an empty (no tree) cell, where $p_g$ is the probability of lightning hitting cell $g$, and $s_g$ is the indicator that is 1 when $g$ has a tree and 0 otherwise.
The denominator is the fraction of the grid that is empty.
The intuition behind this measure is that when fire breaks (i.e., empty cells) lie largely in regions with a high probability of lightning, $C$ will be much larger than 1, whereas if empty cells are distributed uniformly on the grid, $E[C] = 1$.\footnote{These are both formally shown in the supporting online material.}
Figure~\ref{F:fire_breaks} (left) confirms our hypothesis: initially, $C$ is quite high, but as the number of players increases, $C$ approaches 1.
Interestingly, when the number of players is very large ($m=4096$) this result reverses, with $C$ jumping abruptly.
To understand this phenomenon, note that when $m=4096$, each player controls only a $2 \times 2$ subgrid, which is simply too small for a local fire break to be worthwhile unless the fire risk is very high.
Thus, the only players with any incentive to build fire breaks are those close to the epicenter of lightning.

\begin{figure}[ht]
\centering
\begin{tabular}{cc}
\includegraphics[width=\figww]{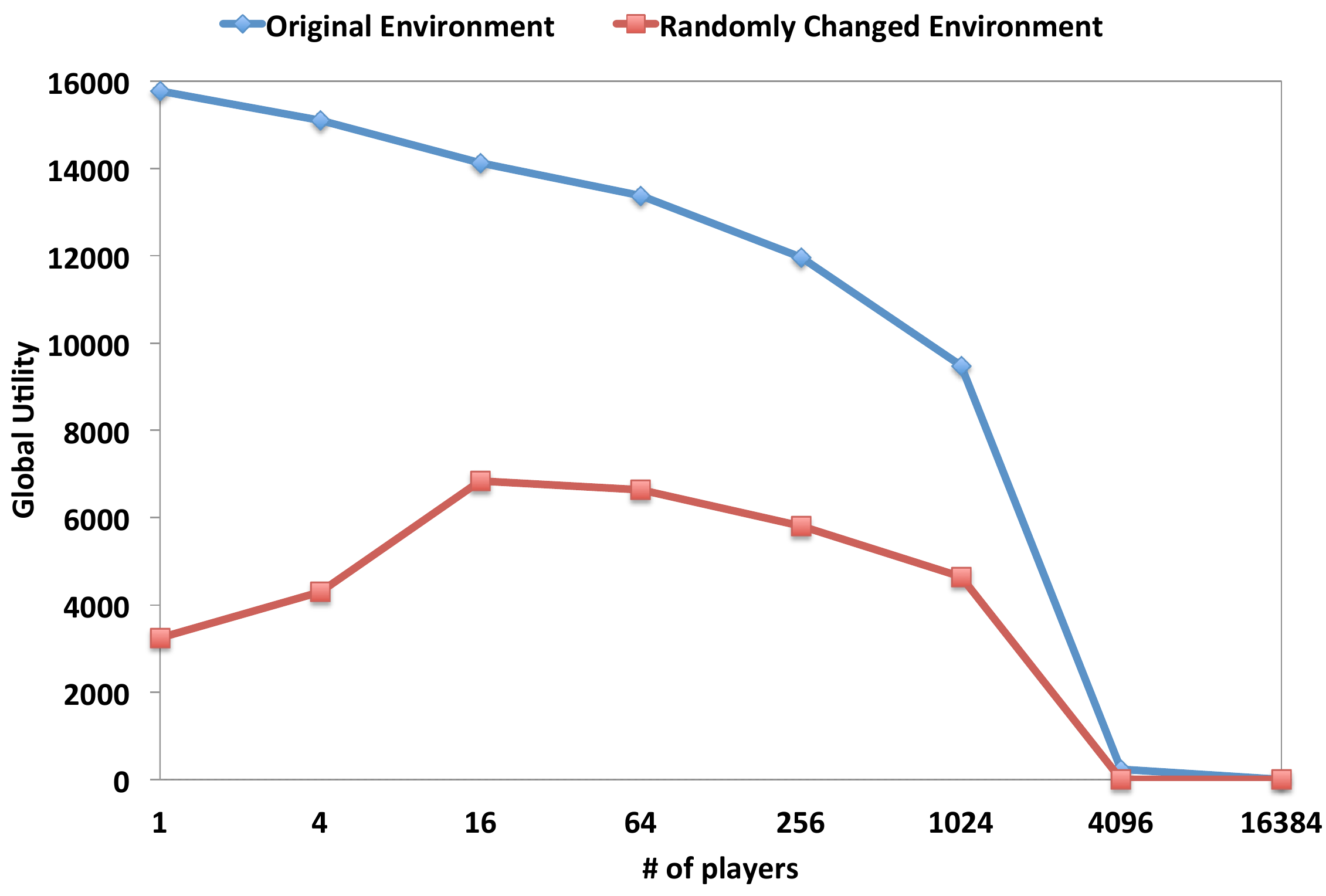} &
\includegraphics[width=\figww]{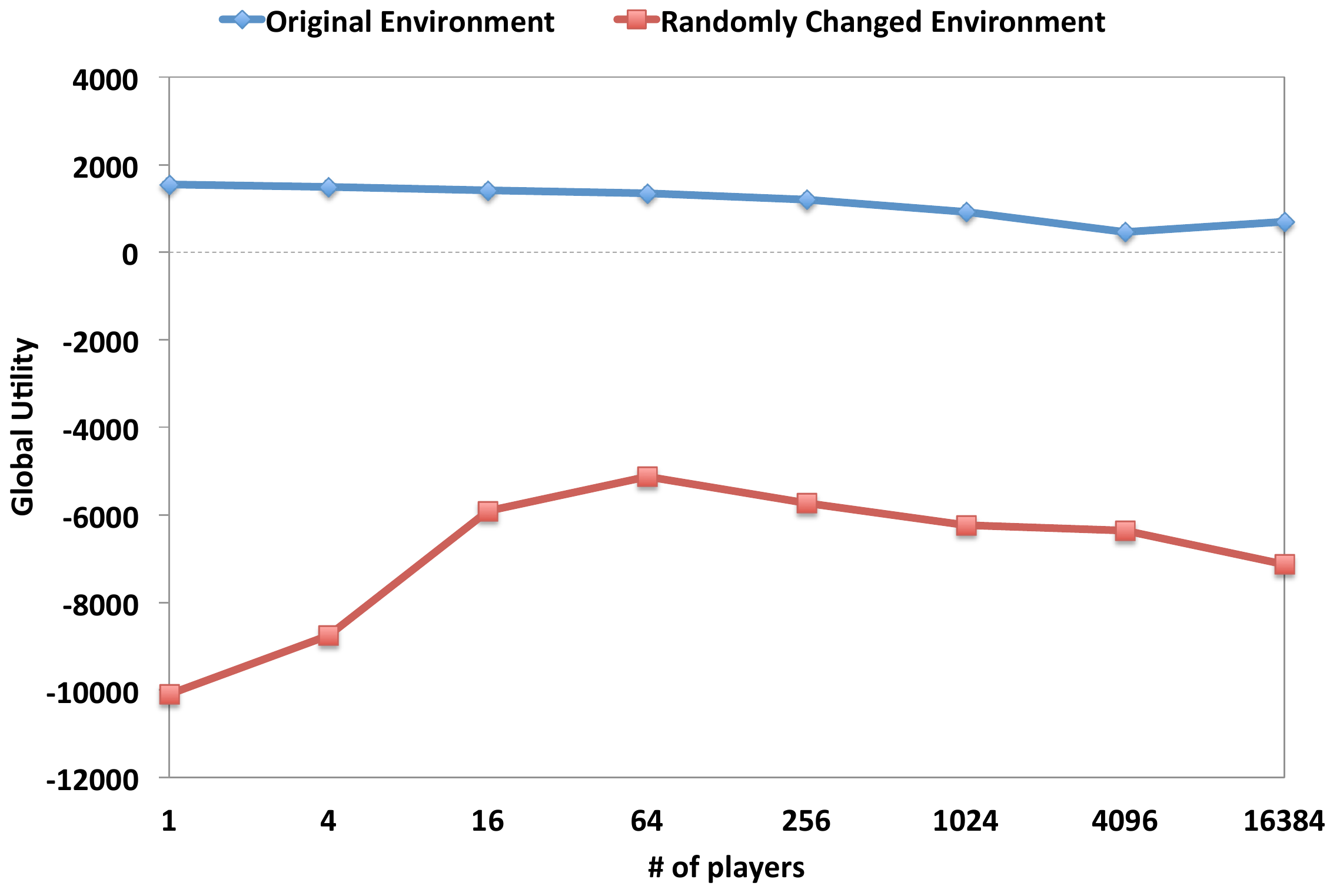}
\end{tabular}
\caption{Fragility of NOT configurations.  Given the (approximate) equilibrium configurations generated for a lightning distribution centered at the upper left corner of the grid, we changed the lightning distribution by generating the center of the Gaussian uniformly randomly from all grid locations.  We the evaluated expected global utility given the altered lightning distribution.  The graph plots averages of repeating this process 30-80 times, as compared to global utility for the original environment. Left: $c=0$.  Right: $c=0.9$.}
\label{F:fragility}
\end{figure}

Considering the spatial distribution of empty grid cells apart from lightning strikes, we see in Figure~\ref{F:fire_breaks} (right) that the centroid of the empty cells begins near the $(0,0)$ point, but approaches the center of the grid with increasing number of players.\footnote{Here again we see that the center shifts back to near the $(0,0)$ point when $m=N/4$, for the same reasons we just outlined.}
Interestingly, even for a moderate number of players ($m=16$), the distribution of fire breaks is nearly homogeneous and almost unrelated to the lightning distribution.
This suggests that global utility would remain relatively robust to changes in the lightning distribution compared to the HOT model.
To verify this, we show in Figure~\ref{F:fragility} average global utility of equilibrium configuration \emph{after the lightning distribution is randomly changed}.
Whether the cost of planting trees is high or low, the figure shows significantly reduced fragility for an intermediate number of players (between 16 and 1024).
Indeed, when cost is high, the system remains less fragile than HOT even in the limiting case of $m=N$.
If we now recall that global utility remains relatively close to optimal across a wide range of settings when $m$ is below 256, our results suggest that the regime of intermediate numbers of players retains the robustness of HOT, while developing some features of SOC that make it less fragile to changes in the environment.

\subsection{Distribution of Burnout Cascades}

One of the central results of both SOC and HOT models is a power-law distribution of burnout cascades.
Since our model generalizes HOT, we should certainly expect to find this power-law distribution in the corresponding special case of $m=1$, at least approximately (since the power-law result in HOT is asymptotic and presumes exact, not approximate, optimization).
We therefore study in some detail how the burnout distribution behaves with respect to the parameters of interest.

\begin{figure}[ht]
\centering
\begin{tabular}{cc}
\includegraphics[width=\figww]{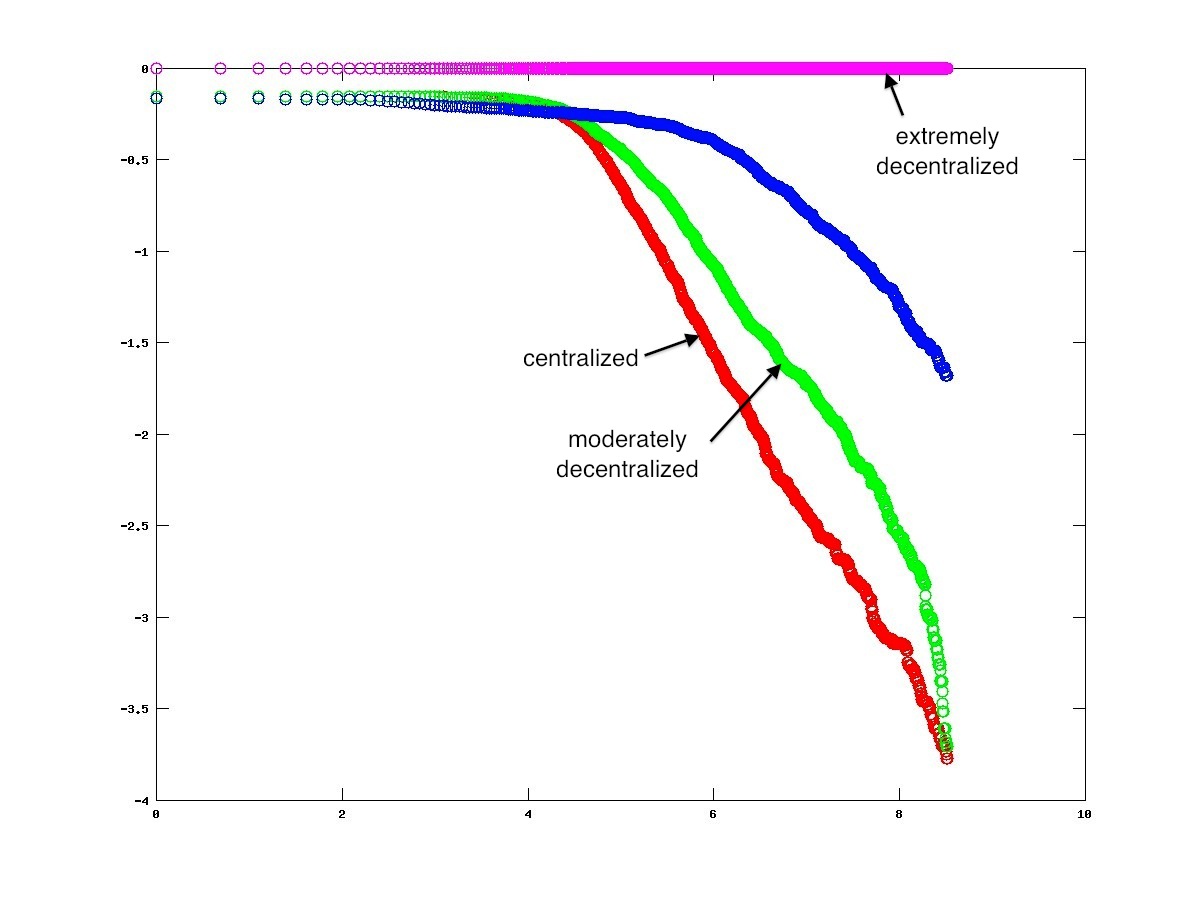}
&
\includegraphics[width=\figww]{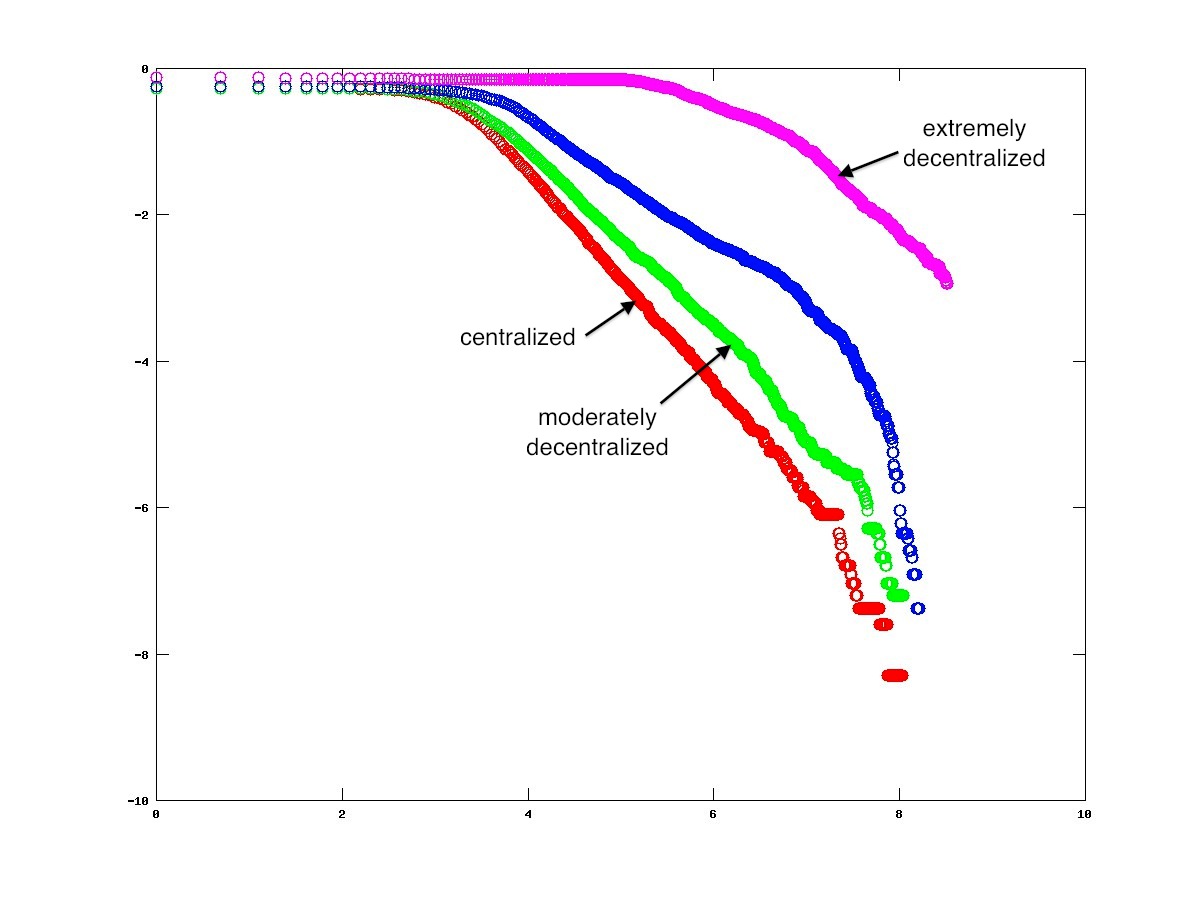} 
\end{tabular}
\caption{Distribution of tree burnout cascades, shown on a log-log plot with $\Pr\{X \geq x\}$ on the vertical axis and $x$ on the horizontal axis, where $X$ is the random variable representing cascade size.  The plots feature (bottom to top) $m = 1$ (red), $m=16$ (green), $m=256$ (blue), and $m=4096$ (purple), with the left plot corresponding to $c=0$ and the right plot corresponding to $c=0.9$.  Both plots correspond to $v=10$.}
\label{F:cascades_v10}
\end{figure}

Figure~\ref{F:cascades_v10} shows fire cascade distributions on the usual log-log plot for $v=10$.
First, when $m=1$ (red points), the results align with the expectation of a near straight line (near power-law distribution) across a range of scales.
Additionally, even when $m$ is greater than 1 but relatively small (green points), the distribution appears linear across a range of scales, suggesting that the power law is likely not unique to the HOT setting.
Once the number of players is large, however, the distribution of cascades less resembles a power law, and begins to feature considerable curvature even in the intermediate scales.
In that sense, the NOT setting with many players is unlike both HOT and SOC\@.

\begin{figure}[ht]
\centering
\begin{tabular}{cc}
\includegraphics[width=\figww]{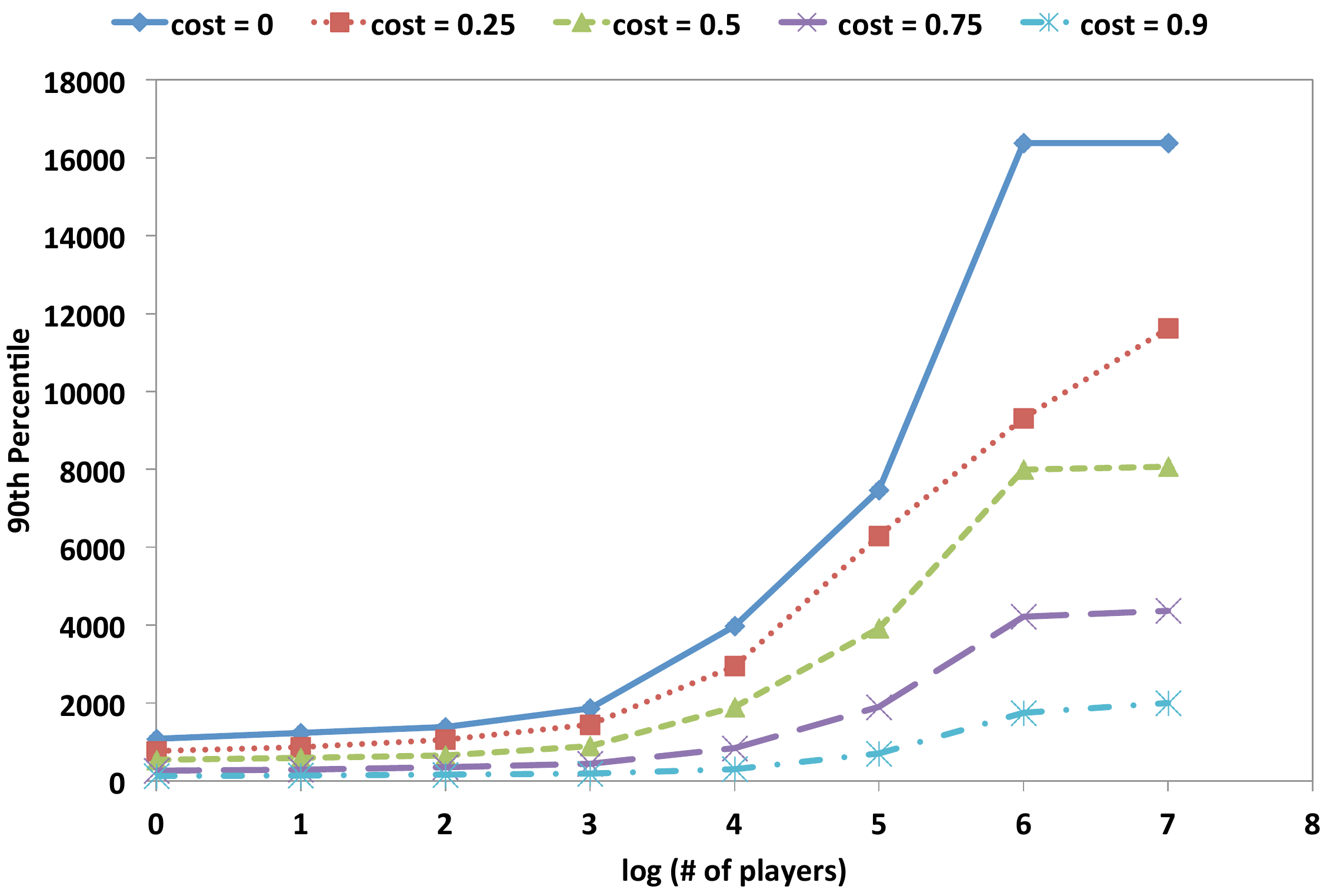}
&
\includegraphics[width=\figww]{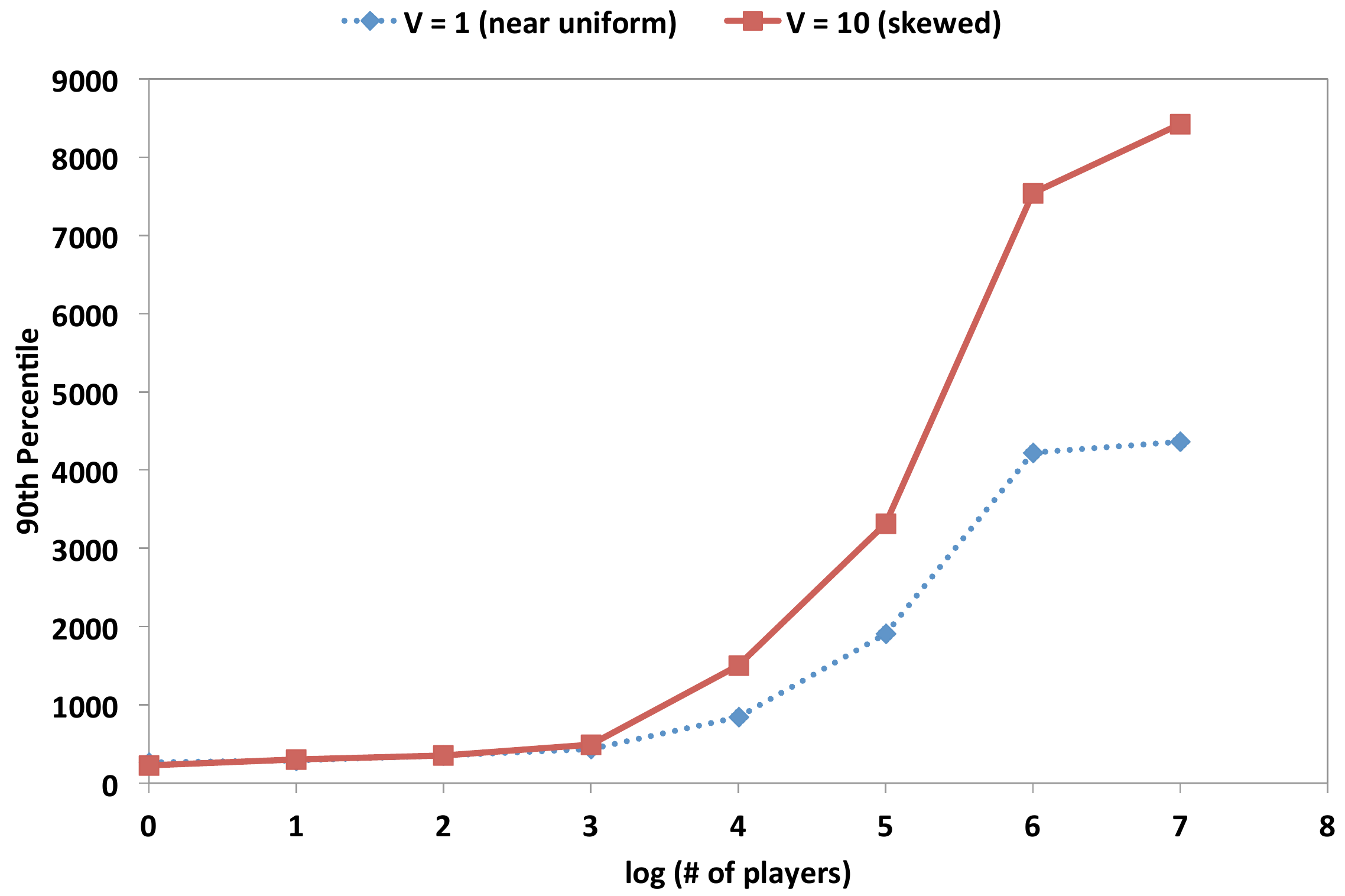} 
\\
\end{tabular}
\caption{Left: 90th percentile as a function of $m$ (plotted on a log scale) for varying values of cost $c$, with $v=1$.  Right: 90th percentile for $c=0.75$ with $v=10$ (top, red) and $v=1$ (bottom, blue).}
\label{F:cascade_quantiles}
\end{figure}

The most important aspect of the cascade distributions is that the tails are systematically increasing with the number of players in all observed settings (this remains the case for Gaussians with greater and smaller variance, not shown here).
We study this in detail in Figure~\ref{F:cascade_quantiles}, which shows the 90th percentile of the burnout distribution as a function of the number of players $m$ for varying cost and variance of the lightning distribution.
The 90th percentile consistently increases with the number of players, confirming the phenomenon of heavier tails with more players that we already observed.
As is intuitive, increasing either the cost of planting trees or the variance of the lightning distribution has a dampening effect on burnout tails: in both cases, more fire breaks are constructed, making very large cascades less likely.

\section{Discussion}

The results described in the previous section show features of both HOT and SOC\@.
When the number of players is small, the NOT setting closely resembles HOT, and, indeed, HOT is a special case when there is a single player.
Perhaps surprisingly, features of HOT persist even when the number of players becomes larger, but as the number of players increases, we also begin to observe many features identified with SOC\@.
The system retains its robustness to the lightning strikes---a key feature of HOT---even when the number of players is relatively large.
It achieves this robustness in part due to the emergence of cooperation between neighboring players, who jointly build fire breaks spanning several players' territories.
The cooperation required to retain near-optimal performance becomes increasingly difficult, however, as the system becomes highly fractured among small domains of influence.

As cooperation becomes less effective, players fall back on protecting their own domain of influence by surrounding it (or parts of it) with deforested land, so long as the fraction of land covered by trees is large enough to make this endeavor worthwhile.
This gives rise to the counterintuitive result that the density of trees initially falls as the number of players increases.

Since even a moderately fractured landscape requires each player to focus on protecting his or her own domain, we observe decreasing correlation between locations of frequent lightning strikes and locations of fire breaks.
With increasing number of players, this correlation systematically decreases, and the spatial distribution of empty cells becomes increasingly homogeneous---striking features of SOC that emerge even when the number of players is not very large and the global performance is still highly robust to lightning strikes.
Thus, the intermediate range of players appears to exhibit both the robustness of HOT and the lack of fragility to changes in the lightning distribution associated with SOC\@.

Another feature of SOC in contrast to HOT is a heavier-tailed distribution of burnout cascades.
We in fact observe that the tail of the burnout distribution becomes heavier with increasing number of players, superficially appearing to shift to an SOC regime.
However, these distributions begin to substantially deviate from a power law even visually, and the setting is therefore in that respect entirely unlike the criticality observed in SOC\@.

\section*{Acknowledgements}
Sandia is a multiprogram laboratory operated by Sandia Corporation, a wholly owned subsidiary of Lockheed Martin Corporation, for the U.S. Department of Energy under contract DE-AC04-94AL85000.

\bibliographystyle{plain}
\bibliography{hotornot}

\appendix
\section{Supporting Online Material}


\subsection{Characterization of $1$- and $N$-player Settings in the 1-D Case}

We begin the analysis by considering the two extremes, $m=1$ and $m=N$, in a simpler model where the forest fire grid is one-dimensional (i.e., a line) and the lightning distribution is uniform.
This analysis will provide some initial findings and intuition that we then carry over into the more complex two-dimensional case.

Without loss of generality, let $k$ be the length of a sequence of planted cells (1's) followed by $l$ unplanted cells (0's) and suppose that $1 \ll k \ll N$.

First, consider the case with $m=1$ and assume that $c < 1 - 1/N$.
Assume that $k$ is identical for all sequences of 1's (when $k \ll N$, this is almost with no loss of generality, since 1's can be swapped, keeping the density constant, without changing the utility) and note that in an optimal solution $l=1$.
The utility of the player (and global utility) is then
\[
u_i(k) = W(k) = \sum_{g \in G} (\Pr_{f}\{\,g = 1 \mid s\,\} - c)s_{g} = N\rho(k)\left(1 - \frac{k}{N} - c\right),
\]
where $\rho(k) = k/(k+1)$.
This function is concave in $k$.
To see this rewrite $u_i$ as
\[
u_i(k) = \frac{Nk(1 - c) - k^2}{k+1}.
\]
Taking the first derivative, we get
\[
u_i' = \frac{N(1-c) - k^2 - 2k}{(k+1)^2}.
\]
Differentiating again we get
\[
u_i'' = -\frac{2(1 + N(1-c))}{(k+1)^3} < 0,
\]
and, hence, $u_i$ is concave in $k$.

Thus, treating $k$ as a continuous variable, which is approximately correct when $k \gg 1$, the first-order condition gives us the necessary and sufficient condition for the optimal $k^*$.
This condition is equivalent to
\[
k^2 + 2k - N(1-c) = 0.
\]
The solutions to this quadratic equation are
\[
k = \frac{-2 \pm \sqrt{4 + 4N(1-c)}}{2}.
\]
Since $k$ must be positive, we can discard one of the solutions, leaving us with
\[
k^* = \sqrt{N(1-c) + 1} - 1.
\]

Evaluating $\rho$ and $W$ at $k^*$, we get
\[
\rho(k^*) = \frac{\sqrt{N(1-c) + 1} - 1}{\sqrt{N(1-c) + 1}}
\]
and
\[
u_i(k^*) = W(k^*) = \rho(k^*) (N(1-c) - \sqrt{N(1-c) + 1} - 1).
\]
We can observe that $\rho(k^*)$ tends to 1 as $N$ grows, while $W(k^*)$ tends to $N(1-c)$  (all derivations are shown in the supporting online material).

Consider next the case with $m=N$.
While there are many equilibria, we can precisely characterize upper and lower bounds on $k$ and $l$, and, consequently, the set of equilibria.
First, we note that $l$ must be either 1 or 2; otherwise, by the assumption that $c < 1 - 1/N$, the player governing any grid cell that is not adjacent to a sequence of 1's will prefer to plant a tree.
Formally, we first note that by definition, $l > 0$.
Suppose $l > 2$ and, thus, there is a player not planting a tree who is not adjacent to another with $s_g = 1$.
Then his utility from planting is $1 - 1/N - c$, and he (weakly) prefers not to plant as long as $1 - 1/N -c \leq 0$ or $c \geq 1 - 1/N$, which is ruled out by our assumption that $c < 1 - 1/N$.

Second, we can get an upper bound on $k$ by considering the incentive of a player that is part of the sequence of 1's.
This player will prefer to plant as long as $1 - k/N - c \geq 0$, giving us $k^E \leq N(1-c)$.
A well-known measure of the impact of equilibrium behavior on global utility is the ``price of anarchy'', the ratio of optimal global utility, here $W(k^*)$, to global utility at the worst-case equilibrium~\cite{Koutsoupias99,Roughgarden05,Nisan07}.
The upper bound on $k^E$ gives us the worst-case equilibrium from the perspective of global utility, with $W(k^E) = 0$ resulting in an infinite price of anarchy (that is, global utility in the worst-case equilibrium is arbitrarily worse than optimal for a large enough number of players and grid cells $N$).

Looking now at the lower bound on $k^E$, we can distinguish two cases, $l=1$ and $l=2$.
When $l=2$, either player not planting a tree prefers not to plant as long as $1 - (k+1)/N - c \leq 0$, and, therefore, $k^E \geq N(1-c)-1$.
For $l=1$, suppose that the two sequences of 1's on either side of the non-planting player have lengths $k$ and $k'$.
The player will prefer not to plant as long as $1 - (k + k' +1)/N - c \leq 0$, where we are adding $k$ and $k'$ since he will be joining the two sequences together if he plants.
This gives us $k + k' \geq N(1-c) - 1$.
Since we are after a lower bound, suppose without loss of generality that $k \leq k'$.
We then get $k^E \geq [N(1-c) - 1]/2$.
It is instructive to apply now another measure of the impact of equilibrium behavior, the ``price of stability'', defined as the ratio of optimal global utility to global utility at the \emph{best-case} equilibrium~\cite{Nisan07,Anshelevich08}.
The best-case equilibrium in our case has $l=1$ and $k^E = [N(1-c) - 1]/2$, and the asymptotic price of stability is 2.

Now we compare the density at equilibrium and at the optimal configuration.
We are looking for the conditions under which the equilibrium density is strictly higher.
Notice that it certainly isn't always the case.
For example, if $c > 1 - 1/N$, no trees will be planted at all in equilibrium or in an optimal configuration.
Consequently, the density will be 0 in both cases.
When $N(1-c) \gg 1$ and $N$ is large, the density in the best-case equilibrium is
\[
\rho(k^E) = \frac{N(1-c) - 1}{N(1-c) + 1}.
\]
Thus, $\rho(k^E) > \rho(k^*)$ iff
\begin{eqnarray*}
\frac{N(1-c) - 1}{N(1-c) + 1} &>&  \frac{\sqrt{N(1-c) + 1} - 1}{\sqrt{N(1-c) + 1}}\\
\Leftrightarrow (N(1-c) - 1)\sqrt{N(1-c)+1} &>& (N(1-c)  + 1) (\sqrt{N(1-c)+1} - 1)\\
\Leftrightarrow 2\sqrt{N(1-c)+1} &<& N(1-c) + 1\\
\Leftrightarrow 4(N(1-c)+1) &<& (N(1-c))^2 + 2N(1-c) + 1\\
\Leftrightarrow 2N(1-c) + 3 &<& (N(1-c))^2.
\end{eqnarray*}
Solving the corresponding quadratic inequality gives us the condition that
\[
N(1-c) > 3.
\]
Since we assume $N(1-c) \gg 1$ throughout, we effectively have that $\rho(k^E) > \rho(k^*)$ under the assumptions operational here.

\subsection{Equilibria When $c=0$ and $m=N$}

Suppose that each player controls a single grid cell, i.e., $m=N$.
When cost of planting trees is 0, there are only two Nash equilibria: one with every player planting a tree, and another with a single player not planting.
Indeed, planting is a weakly dominant strategy for every player.
To see this, suppose that the number of players planting is $z < N-1$, and consider a player who is not planting a tree.
If he decides to plant, the probability of his tree burning down is at most $(N-1)/N < 1$, and so the player has a strict incentive to plant.
Furthermore, since there is no cost of planting, any player who is planting a tree does not lose anything by doing so.
Thus, every player strictly prefers to plant as long as $z < N-1$, and weakly prefers to plant when $z=N-1$ (in which case expected utility is zero whether he plants or not).
Finally, every player planting is clearly an equilibrium, and the only other equilibrium has a single player who does not plant (since he is indifferent, and every other player strictly prefers to plant if that player does not).

\subsection{Details of Equilibrium Approximation}

We now present the details of the algorithms we used to approximate equilibria.
First, we show the ``outer loop'' algorithm for best response dynamics as Algorithm~\ref{A:br}.
\begin{algorithm}
\caption{BestResponseDynamics($T_{br}$, $p_{player}$)}
\label{A:br}
\begin{algorithmic}
\STATE $s_g \leftarrow 0 \ \forall g \in G$
\FOR{$n=1$ to $T_{br}$}
  \FOR{$i=1$ to $m$}
    \STATE Fix $s_{-i}$
    \IF{RAND $\leq p_{player}$}
       \STATE $\hat{s}_i \leftarrow \mathrm{OPT}(s_{-i})$
    \ELSE
       \STATE $\hat{s}_i \leftarrow s_i$
    \ENDIF
    \STATE $s_i \leftarrow \hat{s}_i$
  \ENDFOR
\ENDFOR
\end{algorithmic}
\end{algorithm}
The parameter $T_{br}$ varies depending on the number of players.
For example, if there is just one player, $T_{br} = 1$, whereas $T_{br} = 50$ when $m = N$.
The variation is a consequence of extensive experimentation looking at sensitivity of results to increasing the number of iterations.
Our values are high enough that results do not change appreciably when the number of iterations increases.
We set $p_{player} = 0.9$.

For each player selected by the random biased coin flip (``RAND'' is a uniform random number on the unit interval), the algorithm calls OPT() to approximate the best response of the player to a fixed grid configuration chosen by the others.
Our choice for this procedure is sampled fictitious play, which is shown in pseudocode as Algorithm~\ref{A:opt}.
\begin{algorithm}
\caption{OPT($s_{-i}, T_{opt}, p_{cell}, \alpha, h$)}
\label{A:opt}
\begin{algorithmic}
\STATE $s_g \leftarrow 0 \ \forall g \in G_i$
\STATE $H \leftarrow ()$ \quad \//\//\ Initialize history of past choices $H$ to an empty list
\FOR{$n=1$ to $T_{opt}$}
  \STATE $s'_i \leftarrow \mathrm{ChooseActions}(i, \alpha, H)$
  \STATE $\hat{s}_i \leftarrow s_i$
\FOR{$g \in G_i$}
     \IF{RAND $\leq p_{cell}$ OR $|G_i| = 1$}
       \IF{$u_i(s_g=1,s'_i,s_{-i}) > u_i(s_g = 0,s'_i,s_{-i})$}  
         \STATE $\hat{s}_g \leftarrow 1$
       \ELSE
         \STATE $\hat{s}_g \leftarrow 0$
       \ENDIF
    \ENDIF
  \ENDFOR
  \STATE $\mathrm{append\_back}(H,\hat{s}_i)$ \quad \//\//\ Add $\hat{s}_i$ at the end of list $H$
  \IF{$|H| > h$}
     \STATE $\mathrm{remove\_front}(H)$ \quad \//\//\ Remove the first element
  \ENDIF
  \IF{$u_i(\hat{s}_i,s_{-i}) > u_i(s_i,s_{-i})$}
     \STATE $s_i \leftarrow \hat{s}_i$
  \ENDIF
\ENDFOR
\RETURN $s_i$
\end{algorithmic}
\end{algorithm}
Here, RAND() when called with a list argument picks a uniformly random element of the list.
$u_i()$ is a call to an oracle (a simulator) to determine $i$'s utility in a particular grid configuration.
$u_i(s_g = a, s'_i, s_{-i})$ denotes utility when $i$ plays according to $s'_i$, except he sets $s_g = a$.
We set history size $h = 1$ and exploration parameter $\alpha = 0$.
Thus, each grid cell at iteration $t$ is always best-responding to the grid configuration from iteration $t-1$.
We set $p_{cell} = \max\{0.05,1/N_i\}$.
Thus, on average, one player best-responds in each iteration.
Our parameters for both the optimization routine and the best response routine were chosen based on extensive experimentation.

Algorithm~\ref{A:opt} uses the subroutine ChooseActions(), which is specified as Algorithm~\ref{A:chooseActions}.
\begin{algorithm}
\caption{ChooseActions($i, \alpha, H$)}
\label{A:chooseActions}
\begin{algorithmic}
\FOR{$g \in G_i$}
   \IF{RAND $\leq \alpha$ OR $H = ()$}
     \STATE $s_g \leftarrow \mathrm{RAND}((0,1))$
   \ELSE
     \STATE $s_g \leftarrow \mathrm{RAND}(H)_g$
   \ENDIF
\ENDFOR
\RETURN $s_i$
\end{algorithmic}
\end{algorithm}

In Table~\ref{T:num_itr} we specify the number of iterations used for the outer loop (best response dynamics) and inner loop (approximate optimization).
\begin{table}
\centering
\begin{tabular}{|r|r|r|}\hline
\# players & $T_{br}$ & $T_{opt}$ \\ \hline
1 & 1 & 200\\ \hline
4 & 5 & 120 \\ \hline
16 & 20 & 80 \\ \hline
64 & 20 & 80\\ \hline
256 & 20 & 80\\ \hline
1024 & 40 & 80\\ \hline
4096 & 20 & 35\\ \hline
16384 & 50 & 1\\ \hline
\end{tabular}
\caption{Numbers of iterations of best response dynamics and sampled fictitious play in the 2nd and 3rd column respectively.}
\label{T:num_itr}
\end{table}

\subsection{Relationship Between Lightning Distribution and Empty Cells}

Recall our measure that captures the relationship between the distribution of empty cells (fire breaks) on the grid and the lightning distribution:
\[
C = \frac{\sum_{g \in G} p_g (1-s_g)}{1-\rho}.
\]
We now formally demonstrate that (a) $C > 1$ when empty cells have the largest probability of lightning and (b) $E[\sum_{g \in G} p_g (1-s_g)] = 1 - \rho$ when empty cells are chosen uniformly randomly on the grid.
First, suppose that the cells that have the $L$ highest lightning probabilities on the grid are empty, and let probabilities be ranked from highest to lowest such that $p_l$ indicates $l$th highest probability of lightning on the corresponding grid cell $g_l$.
Further, suppose that no two $p_l$ are the same.
Then
\[
\sum_{g \in G} p_g (1-s_g) = \sum_{l=1}^N p_l (1-s_{g_l}) = \sum_{l=1}^L p_l > \frac{L}{N},
\]
where the last inequality follows since there are no ties between $p_l$.
Since $\frac{L}{N} = 1 - \rho$, the result follows.

Next we show that if empty cells are uniformly distributed, $E[\sum_{g \in G} p_g (1-s_g)] = 1 - \rho$.
First, note that $1-\rho = L/N$ if $L$ cells are empty and the rest have a tree.
Now, suppose that each cell is empty with probability $q = L/N$.
Then
\[
E[\sum_{g \in G} p_g (1-s_g)] = \sum_{g \in G} p_g E[1-s_g] = q \sum_{g \in G} p_g = q.
\]
Since $q = L/N = 1 - \rho$, the result follows.

\subsection{Tree Fines}

\begin{figure}[ht]
\centering
\begin{tabular}{cc}
\includegraphics[width=\figww]{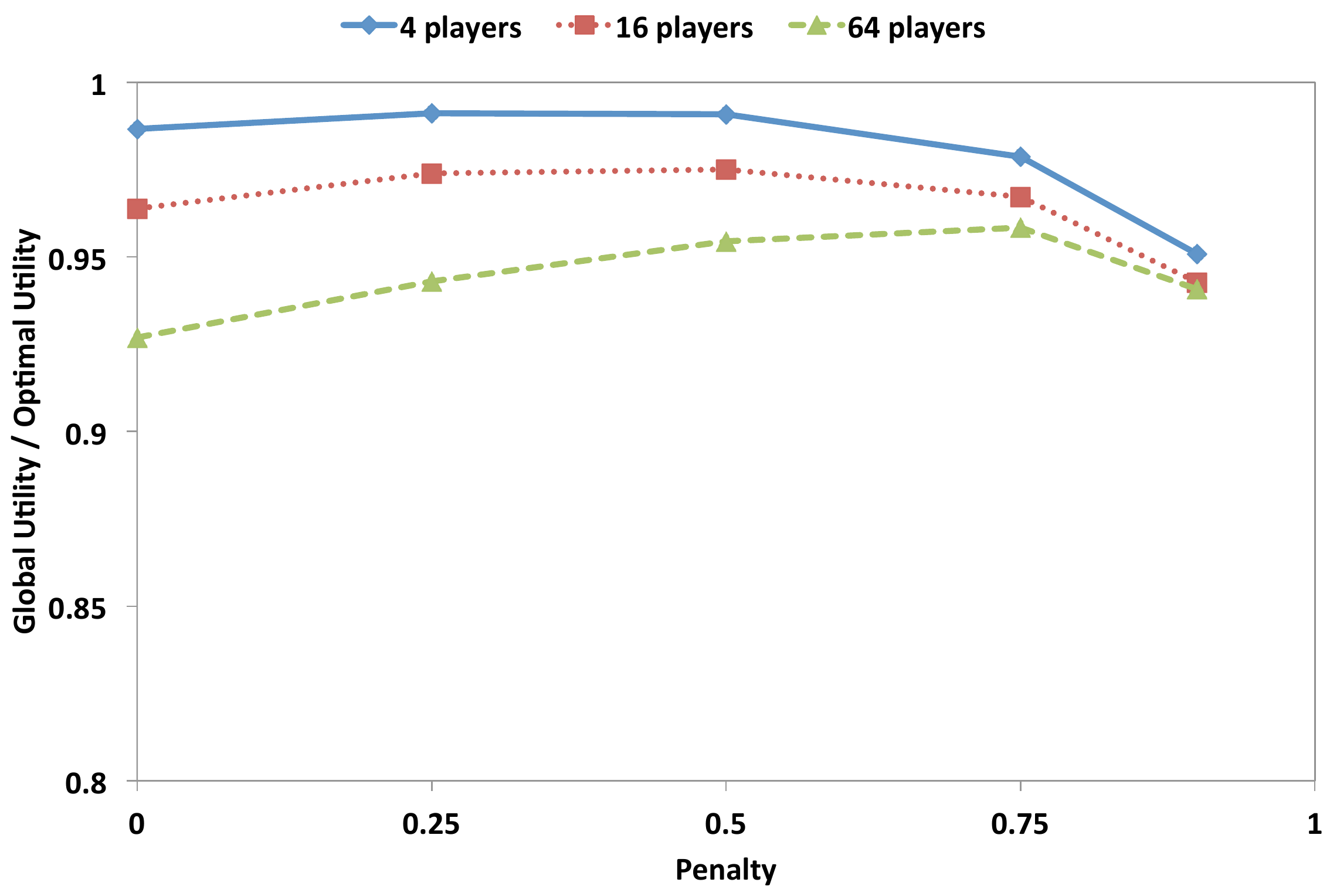}
&
\includegraphics[width=\figww]{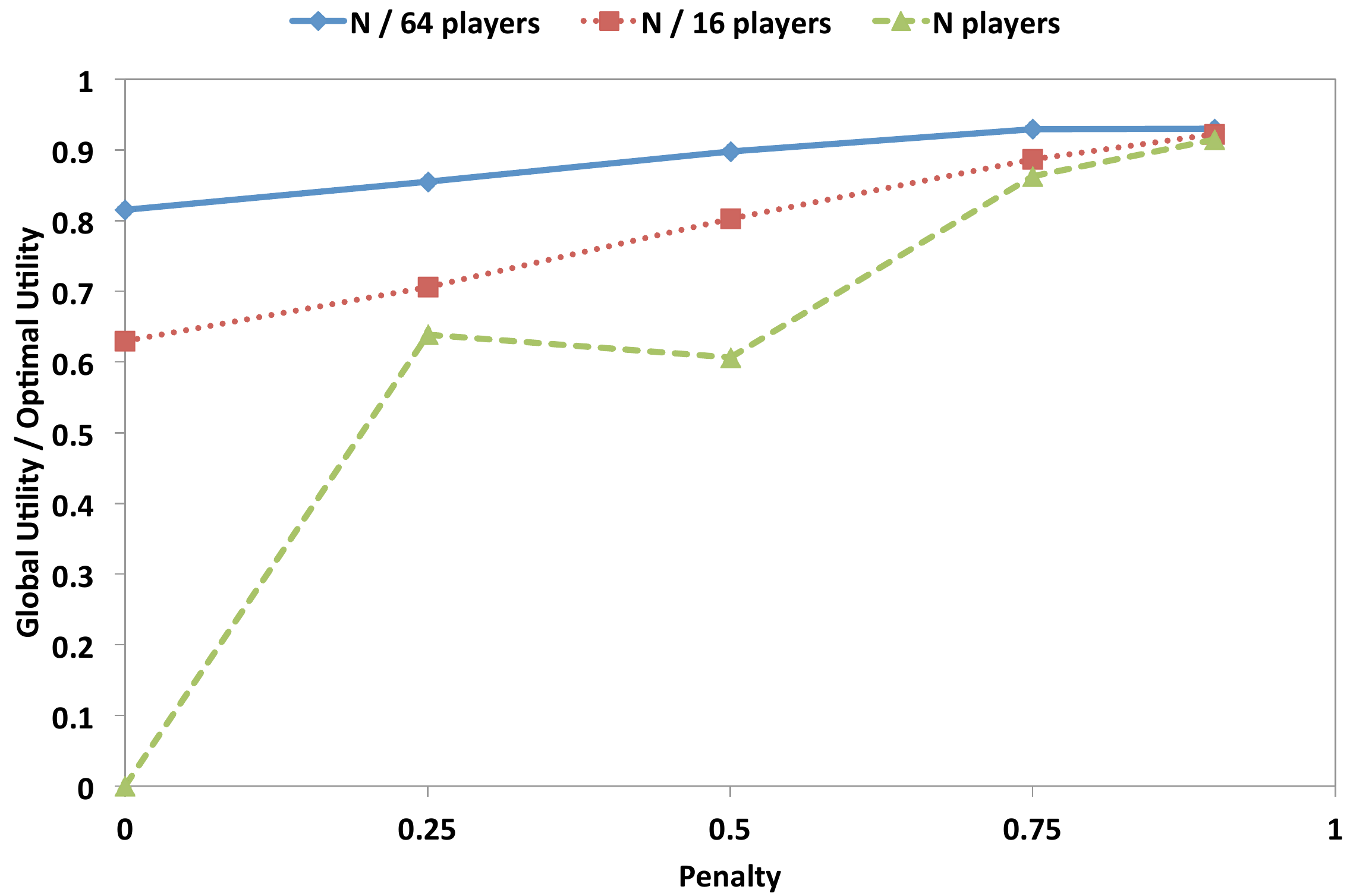} 
\\
\end{tabular}
\caption{Variation of global utility $W(p)$ relative to its optimal value, as a function of penalty amount $p$, with actual cost of planting a tree $c=0$.  Left: ``few'' players, that is, $m\in\{4,16,64\}$.  Right: ``many'' players, that is, $m\in\{N/64,N/16,N\}$.}
\label{F:planting_fines}
\end{figure}

Another interesting inquiry concerns the question of policy: can imposing a fine on planting trees alleviate the impact of negative externalities when the number of players is large?
To this end, suppose that $p$ is a penalty for planting a tree.
Global utility is then redefined as
\[
W(p) = Y(p) - cN\rho(p),
\]
where $Y(p)$ and $\rho(p)$ are the equilibrium global yield and density respectively when the true cost of planting a tree is $c$ but each player perceives it to be $c + p$.
As Figure~\ref{F:planting_fines} suggests, the results are somewhat mixed.
First, considering just the plot on the right, we note that a small penalty can have a large impact when $m=N$.
Increasing the penalty further, however, seems to improve global utility only slightly when the number of players is large.
On the other hand, the plot on the left suggests that when the number of players is small, increasing player costs is at best ineffective, and at worst may actually lower global utility.
In either case, outcomes never quite reach the optimum, although they come quite close when the number of players is small.
The simple policy of raising costs of players via fines is therefore a relatively ineffective instrument here, and can at times be counterproductive.

\end{document}